\documentclass[11pt,a4paper,tightenlines]{revtex4}

\usepackage{color} 

\usepackage{graphicx}        

\usepackage{natbib}         

\usepackage{amssymb}        

\usepackage{url}             

\newcommand{\bK}{ {\mathbf K } }

\newcommand{\K}{ {\rm K } }

\newcommand{\bk}{ \mbox{\boldmath$k$} }

\newcommand{\cn}{ {\cal N} }

\newcommand{\cT}{ {\cal T} }

\newcommand{\Cref}{ {C_{\rm ref}} }

\newcommand{\br}{{\mbox{\boldmath$r$}}}

\newcommand{\bbv}{ \mbox{\boldmath${v}$} }
\newcommand{\bbu}{ \mbox{\boldmath${u}$} }

\newcommand{\bnabla}{ \mbox{\boldmath$\nabla$} }

\newcommand{\id}{ {\rm d} }

\newcommand{\cK}{ {\cal K}}

\newcommand{\bcK}{ \mbox{\boldmath$\cal K$} }
\newcommand{\bcT}{ \mbox{\boldmath$\cal T$} }

\newcommand{\oned}{{\mbox{\tiny {\rm 1D}}}}
\newcommand{\twod}{{\mbox{\tiny {\rm 2D}}}}
\newcommand{\tpod}{{\mbox{\tiny {\rm 2+1D}}}}

\newcommand{\trt}{{\mbox{\tiny${\rm tt}$}}}
\newcommand{\sigmaunc}{{\mbox{\tiny{\rm unc}}}}
\newcommand{\sigmacorr}{{\mbox{\tiny{\rm corr}}}}

\newcommand{\uu}{ {u}}
\newcommand{\vv}{ {v}}

\newcommand{\aap}{    {\it Astron. Astrophys.}}

\newcommand{\apjl}{   {\it Astrophys. J. Lett.}}

\newcommand{\apjs}{   {\it Astrophys. J. Suppl.}}

\newcommand{\physscr}{  {\it Physica Scripta}}

\newcommand{\solphys}{{\it Solar Phys.}}

\newcommand{\f}   {\mbox{${\rm f}$}}

\newcommand{\pone}{\mbox{${\rm p}_1$}}

\newcommand{\ptwo}{\mbox{${\rm p}_2$}}

\newcommand{\pthree}{\mbox{${\rm p}_3$}}

\newcommand{\pfour}{\mbox{${\rm p}_4$}}

\newcommand{\inlinecite}[1]{\cite{#1}}
\newcommand{\opencite}[1]{\cite{#1}}

\begin{document}

\title{High-resolution mapping of flows in the solar interior: Fully consistent OLA inversion of helioseismic travel times}

\author{J. Jackiewicz}
\author{L. Gizon}
\affiliation{Max-Planck-Institut f\"{u}r Sonnensystemforschung, 37191 Katlenburg-Lindau, Germany}
\author{A.~C. Birch}
\affiliation{NWRA CoRA Division, Boulder, CO, 80301}





\begin{abstract}

To recover the flow information encoded in travel-time data of time-distance helioseismology, accurate forward modeling and a robust inversion of the travel times are required. We accomplish this using three-dimensional finite-frequency travel-time sensitivity kernels for flows along with a 2+1 dimensional (2+1D) optimally localized averaging (OLA) inversion scheme. Travel times are measured by ridge filtering  MDI full-disk Doppler data and the corresponding  Born sensitivity kernels are computed for these particular travel times. We also utilize the full noise covariance properties of the travel times which allow us to accurately estimate the errors for all inversions. The whole procedure is thus fully consistent. Due to ridge filtering, the kernel functions separate in the horizontal and vertical directions, motivating our choice of a 2+1D inversion implementation. The inversion procedure also minimizes cross-talk effects among the three flow components, and  the averaging kernels resulting from the inversion show very small amounts of cross-talk. We obtain  three-dimensional maps of vector solar flows in the quiet Sun at spatial resolutions of  $7-10$~Mm using generally $24$~h of data. For all of the flow maps we provide averaging kernels and the noise estimates. We present examples to test the inferred flows, such as a comparison with Doppler data, in which we find a correlation of 0.9.  We also present results for quiet-Sun supergranular flows  at different depths in the upper convection zone. Our estimation of the vertical velocity shows good qualitative agreement with the horizontal vector flows.  We also show vertical flows measured solely from f-mode travel times. In addition, we demonstrate how to directly invert for the horizontal divergence and flow vorticity. We finally study inferred flow-map correlations at different depths and find a rapid decrease in this correlation with depth, consistent with other recent local helioseismic analyses.

\end{abstract}



\maketitle

%


%

\section{Introduction}

\label{sec-introduction}

Time-distance helioseismology \cite{duvall1993} is a set of tools that
measures and interprets  the travel times of seismic waves propagating from
one point on the solar surface to any other point. It has been shown that
these travel times contain information about solar flows \cite[among others]{kosovichev1996,duvall1997,duvall2000,gizon2000,zhao2001}. This paper focuses on the inversion of travel times to obtain high spatial
resolution maps of near-surface vector flows in quiet-Sun regions. What is unique to this study is
that it is the first fully consistent inversion in time-distance
helioseismology. The consistency is described by several factors: (1) we
measure travel times with the same definition with which the travel-time
sensitivity kernels are computed; (2) we use three-dimensional (3D)
finite-frequency Born sensitivity kernels which are necessary to detect
 flow structures that have spatial scales on the order of the mode wavelength - this is the regime where the commonly used ray approximation fails \cite[for example]{birch2004b}; (3) we use the full noise covariance properties of the travel times as an ingredient in the inversion; (4) the inversion procedure we choose to implement is `optimal', in that it simultaneously   achieves the best possible spatial resolution while minimizing the magnification of the errors. Furthermore, the regularization is carried out in both the horizontal and vertical directions.

We have developed a novel two plus one dimensional (2+1D) inversion scheme
based on the well-known subtractive optimally localized averages (SOLA) technique
\cite{pijpers1992,jackiewicz2007a}. The inversion procedure explicitly minimizes the cross-talk effects among the three flow components by imposing constraints on the averaging kernels. An important aspect that we introduce to this procedure is that of ridge-filtered travel-time measurements, whereby only wave packets of a particular radial order are used. Sets of point-to-annulus travel times are then computed for the f, \pone, \ptwo, \pthree, and \pfour\ ridges. This is quite different than the usual phase-speed filtering implemented for travel-time measurements. Subsequently,  the sensitivity kernels are also  computed as corresponding ridge-filtered quantities to match the travel times. The 2+1D inversion is motivated by the
observation that to a very good approximation the 3D Born ridge-filtered sensitivity kernels separate
into the product of a 2D horizontal function and a 1D function in depth.  

Together, these ingredients allow us to infer all three components of the
vector flow in the near-surface layers of the quiet-Sun convection zone. 
In previous work, we determined  for the first time the maximum amplitude of
flows which  can be reliably recovered from a linear model of the travel-time perturbations
\cite{jackiewicz2007b}. The supergranular and other quiet-Sun flows  (with velocities $\le 400$~m\,s$^{-1}$) that we detect in this study  fall within this
range, giving us confidence in the reliability of the method. The horizontal
spatial resolution of the inversion presented here ($\sim 7-10$~Mm, depending on the observation time, depth, etc.) is on the order of, or even in some cases below, the wavelength of the
waves used in the analysis (typically $5-20$~Mm, depending on the dominant modes). In addition, we perform a direct
measurement of the vertical component of the flow (without simply invoking
mass conservation) and with confidence that the cross talk between the horizontal and vertical components has been minimized that minimizes. Interestingly, we find that we can determine the vertical velocity even from f-mode travel times.  We also directly invert for the horizontal divergence of the flow  as well
as the vertical component of the flow vorticity. 

As the main  aim of this paper is to develop the inversion procedure  and to
perform tests of it on real solar data, we postpone the main interpretation of the
results to a future publication.  We typically show quiet-Sun flows that have been obtained from $24$~h of data, in order to maximize the signal-to-noise ratio from the supergranulation signal \cite{gizon2004}. In all of the inferred flow maps we provide estimates of the noise and the spatial resolution.

The paper is organized as follows: In Section~\ref{sec:tt} we describe the data and the ridge-filtered travel-time measurements. That is followed by a brief discussion of the forward modeling, i.e., the computation of sensitivity kernels and the noise covariances that are consistent with the travel times. Since the multi-step inversion procedure is somewhat complicated, we provide an overview in Section~\ref{sec:basic}, followed by two sections that discuss in detail the 2D and 1D parts with plenty of example calculations. Three-dimensional averaging kernels at different depths are presented in Section~\ref{sec:3davekerns}, and finally the various flow maps are presented, tested, and discussed in Section~\ref{sec:results}. We end  with a summary of the results and a discussion of current and future work.

%


%

\section{Data and ridge-filtered travel-time measurements}
\label{sec:tt}

For this study we  use dopplergram data from the Michelson Doppler Imager
\cite{scherrer1995} on board \emph{SOHO} which are full-disk images of $0.12^\circ$
spatial sampling ($\sim 1.46$~Mm) and  $1$~minute  cadence. The region of
interest is an area centered on NOAA AR 9787 observed between 20-28 January
2002 and tracked and remapped courtesy of T.~L. Duvall Jr.\footnote{To
  download these data as well as the corresponding magnetograms and intensity
  images for analysis, visit
  \textsf{http://www.mps.mpg.de/projects/seismo/NA4/DATA/data\_access.html}.}
The size of the full set of velocity data cubes is $512\times512\times1440\times 9$ (two spatial dimensions, $1440$ minutes, $9$ days). We only use the middle seven days for the results shown here. These data are ideal for helioseismic analysis as there is a  sunspot that is large, isolated, and quite stable as it traverses the disk. For our purposes, there are also large regions of relatively quiet Sun in these maps where we will focus our analysis.

We denote a Doppler velocity cube as $\phi(\br,t)$, where 
\begin{equation}
\br=(x,y) 
\end{equation}
is the horizontal coordinate, $x$ and $y$ are the east-west and  north-south directions, respectively, $t$ is time,  and we work in a Cartesian geometry. We filter the
data by  multiplying the Fourier transform of the data cube by  a filter $F_n(\bk,\omega)$ which selects all modes with the same radial order $n$, and removes all others. We call this ridge filtering.  The ridge-filtered data $\Phi_n$ are then given by
\begin{equation}
\Phi_n(\bk,\omega)=F_n(\bk,\omega)\phi(\bk,\omega),
\end{equation}
where $\bk$ is the horizontal wavevector and $\omega$ is the angular
frequency.  The ridges we filter and retain  for this study are the surface-gravity wave
(f-mode) ridge and the first four acoustic (p-mode) ridges. Throughout the text we  carry the index $n$ which   takes the  possible  values   $n=\{\f,\pone,\ptwo,\pthree,\pfour\}$. We refer to these as mode ridges. It is important to note that
this type of filtering is different from the phase-speed filtering that is typically
done prior to any time-distance analysis.

The cross-covariance functions are computed from $\Phi_n(\bk,\omega)$ for three different point-to-annulus geometries, denoted by `oi' (out minus in), `we' (west minus east), and `ns' (north minus south). The `oi' covariances are measured using the wave signal at a given point and the wave signal averaged over a concentric annulus of radius $\Delta$. The `we' (`ns') quantities use the central wave signal along with the wave signal averaged over the annulus but weighted by $\cos\theta$ ($\sin\theta$), where $\theta$ is the angle between the $x$ direction and each point on the annulus. This particular procedure was introduced in \inlinecite{jackiewicz2007a} and is similar to what is usually done in standard time-distance measurements \cite{duvall1997}. The temporal cross-covariance functions are computed for 20 different annulus radii ($\Delta=1.46$~Mm to $29.2$~Mm, incremented by $\id\Delta=1.46$~Mm) for each mode ridge and measurement type, which yields a set of functions $C^{\alpha,n}(\br,t;\Delta)$, where $\alpha={\rm \{oi,we,ns\}}$. Flows introduce asymmetries in time lag $t$ in the cross-covariance functions.

Travel times are then measured according to the procedure developed by \cite{gizon2002,gizon2004}. This method establishes a linear relationship between the travel-time perturbations $\tau$ and the cross covariances, given by

\begin{equation}
\tau^\alpha_n(\br,\Delta) = h_t\sum_t W^n(\Delta,
t)\,\left[C^{\alpha,n}(\br,t;\Delta)-C^{{\rm ref},n}(\br,t;\Delta)\right],
\label{crosscorr}
\end{equation}
where the sum over $t$ denotes a discreet sum over the observation time $T$ (for this work $T=1$~day in most cases),
$h_t=1$~min. is the temporal sampling rate of the data, $W$ are weight
functions, and $C^{\rm ref}$ is a reference cross covariance symmetric in time
lag, which we choose to be the cross covariance computed from our model power
spectrum. This model is tuned  to match the observed power spectrum of each
individual ridge (see Section~\ref{sec:forward}). $C^{\rm ref}$ is  the
inverse Fourier transform of the model power spectrum. Full details about the
$W$ function and equation~(\ref{crosscorr}) can be found in \inlinecite{gizon2004}. The resulting three different types of travel-time measurements are constructed  to have sensitivity to different flow geometries. The `oi' travel times are highly sensitive to horizontal flow divergence, whereas the `we' and `ns' travel times give information about directional flows.

%


%

\section{Forward modeling}
\label{sec:forward}
We now briefly discuss the steps we have taken  to model the ridge-filtered travel-time measurements
described in the previous section, as well as the measurement noise properties. Full details of forward modeling in time-distance helioseismology can be found in \inlinecite{gizon2002}, \inlinecite{gizon2004}, and \inlinecite{birch2007}.

\subsection{Travel-time sensitivity kernels for ridge filtering}

\begin{figure}
\centerline{\includegraphics[width=\textwidth]{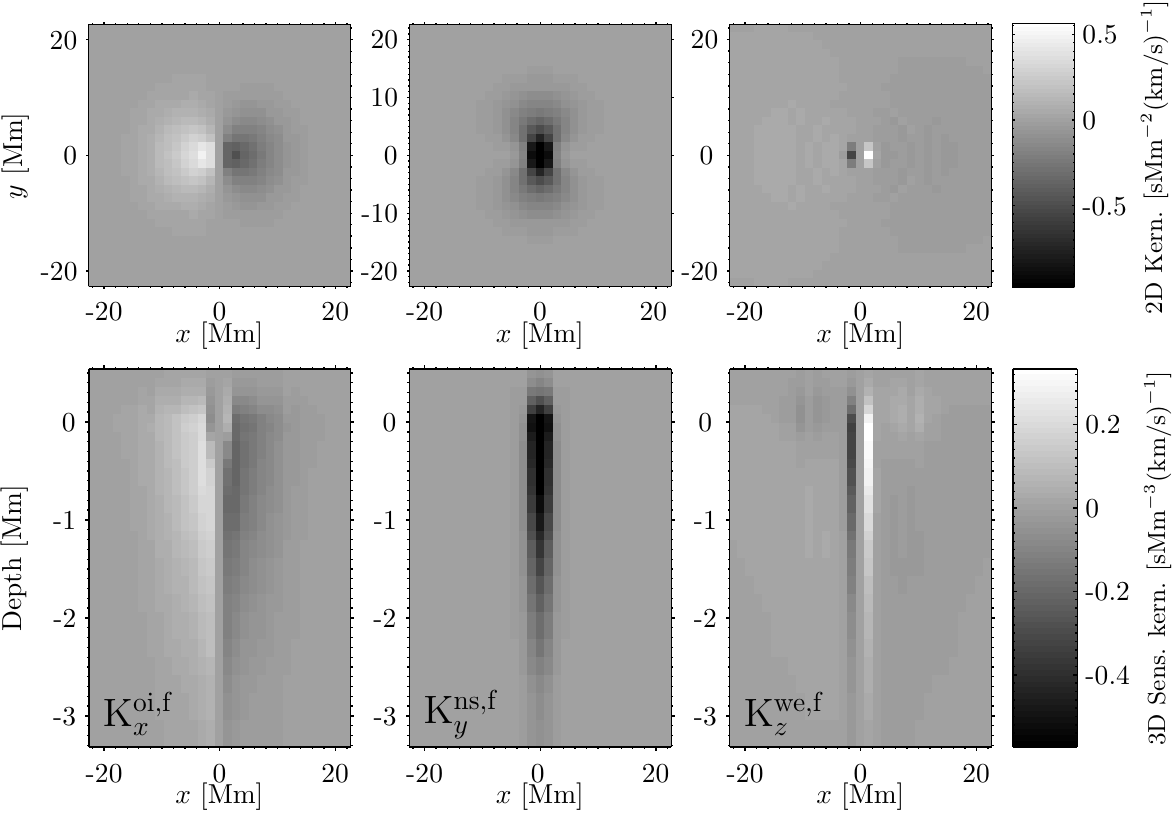}}
\caption{Examples of various sensitivity kernels $\bK^{\alpha,{\rm f}}(\br,z;\Delta)$ used in this study and
    defined in equation~(\ref{tt-k}). They have been obtained by computing weighted azimuthal
    averages of the  point-to-point kernels for the $\alpha$ point-to-annulus geometries. The kernels in the left column give the sensitivity of `oi' f-mode travel times to $\uu_x$, and the kernels in the middle (right) column give the sensitivity of the `ns' (`we') f-mode travel times to $\uu_y$ ($\uu_z$). The top row are 2D kernels after
    integrating the 3D kernels over depth. The bottom row shows depth slices
    along $y=0$ of the 3D kernels.  The type $\alpha$ and mode ridge $n$ of
    the kernels in
    each column is indicated in the bottom panels. For each case,
    $\Delta=10.2$~Mm. The gray scale in the bottom row has been truncated to
    $75$\% of kernel maximum.}
  \label{fig:sens_kerns}
\end{figure}

We consider travel-time measurements of type $\alpha={\rm \{oi,we,ns\}}$ for each distance $\Delta$ and for each mode ridge $n$ as described in the previous section. The travel-time perturbations are related to the small-amplitude flows through a linear relation:

\begin{equation}
\label{tt-k}
\tau_n^\alpha(\br_i;\Delta)=h_r^2h_z\sum_{j,z}\bK^{\alpha,n}(\br_j-\br_i,z;\Delta)\cdot\bbu(\br_j,z)+\cn_\trt^{\alpha,n}(\br_i;\Delta),
\end{equation}
where  $\bK$ denotes the three-dimensional travel-time sensitivity kernel, $\bbu$ is the real
vector flow in the Sun,  $\cn_\trt$ represents  the noise in the travel times (tt), and the sum over
the vertical coordinate $z$ denotes a discreet sum (note $z=0$ at the surface and is negative inside the Sun). The kernels
are computed so that the 
horizontal grid spacing,  $h_x=h_y=1.46$~Mm where  $h_r^2=h_xh_y$,  matches that of  the travel-time
measurements. The vertical grid (with spacing $h_z$) is taken from the
model on which the kernels are computed. The kernel $\bK$ is computed in the first Born
approximation \cite{gizon2002}.  We start with the point-to-point kernels for flows derived
in \inlinecite{birch2007}, computed for the f, \pone, \ptwo, \pthree, and \pfour\ ridges, whose input power spectra have been tuned to
match the ridge-filtered observed power spectra. For each ridge, $20$ kernels are computed, one for each $\Delta$.  The 3D point-to-point kernels are
then  azimuthally averaged according to
the three different point-to-annulus weighting geometries $\alpha$ used for the travel-time
measurements. The total number of point-to-annulus kernels thus obtained is $300$, each kernel having $3$ components. A few examples of f-mode kernels after weighted azimuthal averaging are shown in
Figure~\ref{fig:sens_kerns} for $\br_i=0$. From left to right, the columns show kernels that give the sensitivity to $\uu_x$, $\uu_y$, and $\uu_z$, respectively. The top row shows the resulting 2D kernel
after integration over depth.

\begin{figure}
  \centerline{\includegraphics[width=\linewidth]{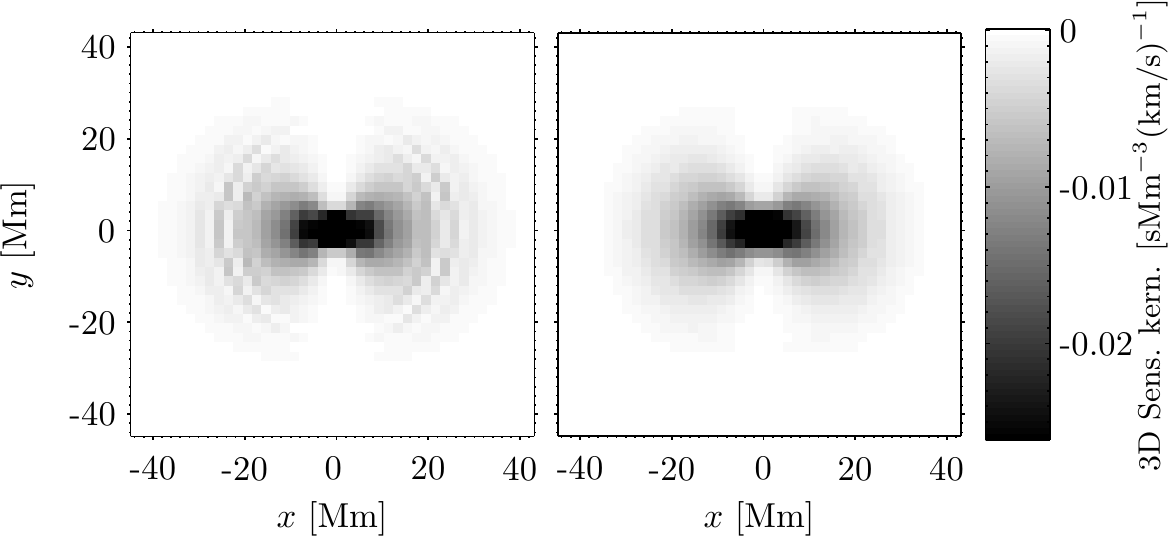}}
   \vspace{-.13\linewidth}
  \centerline{\large \bf    
    \hspace{0.11\textwidth}  \color{black}{(a)}
    \hspace{0.29\textwidth}  \color{black}{(b)} 
    \hfill}
\vspace{1.4cm}
  \centerline{\includegraphics[width=\linewidth]{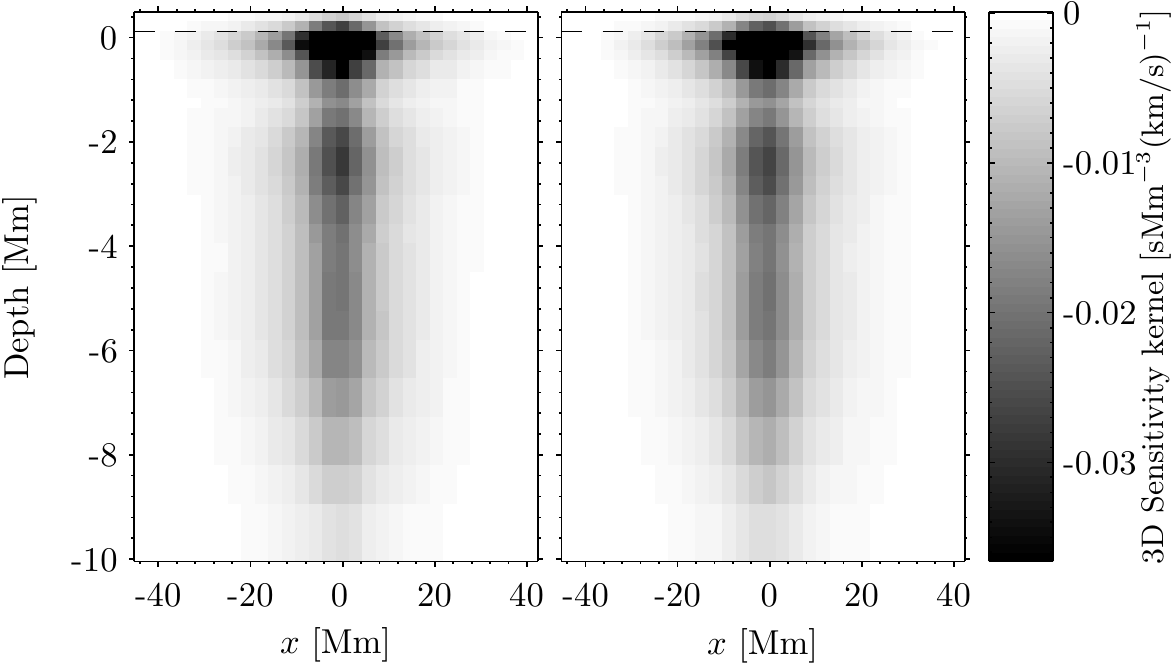}}
  \vspace{-.13\linewidth}
  \centerline{\large \bf    
    \hspace{0.11\textwidth}  \color{black}{(c)}
    \hspace{0.29\textwidth}  \color{black}{(d)} 
    \hfill}
  \vspace{1.5cm}
  \caption{Separability of a sensitivity kernel for \ptwo\ and for ridge filtering. This particular kernel, $\K_x^{\rm we,p_2}$,  gives the sensitivity of `we' \ptwo-mode travel times to $\uu_x$ for $\Delta=23.4$~Mm . (a) Horizontal cut through the kernel at a height of about $100$~km above the photosphere, denoted by the dashed line in the panel below. (b) Horizontal cut through the quantity obtained as the product of the two (normalized) functions, $f(\br)$ and $g(z)$, computed by integrating the kernel in (a) horizontally and over depth, according
    to equations~(\ref{sep1},\ref{sep}).
Panel~(c) is a depth slice
    along $y=0$ of the original kernel from (a), and panel~(d) is a depth slices along $y=0$ of the kernel in (b).  The match between the left and right columns is quite good. The gray scale has been truncated to
    $40$\% of the maximum value for ease of comparison. }
  \label{fig:sep_kerns}
\end{figure}

As mentioned in the introduction,  the motivation behind the  inversion
we choose to perform is that the ridge-filtered kernels to a very good approximation separate into a
horizontal 2D function of $\br$  times a 1D function of depth $z$.  In other
words, dropping the labels $\alpha, n,$ and $\Delta$ for the moment,  the  $i$th component  of the
sensitivity kernel defined in equation~(\ref{tt-k}) may be written as the product
\begin{equation}
\K_i(\br,z)\simeq f_i(\br)g_i(z),
\label{sep1}
\end{equation}
where
\begin{equation}
f_i(\br)=\frac{h_z\sum_z\K_i(\br,z)}{\sqrt{h_r^2h_z\sum_{j,z}\K_i(\br_j,z)}},
\qquad
g_i(z)=\frac{h_r^2\sum_j\K_i(\br_j,z)}{\sqrt{h_r^2h_z\sum_{j,z}\K_i(\br_j,z)}}.
\label{sep}
\end{equation}
In  Figure~\ref{fig:sep_kerns}  we demonstrate the separability for an example \ptwo\ and `we' averaged kernel. Figure~\ref{fig:sep_kerns}a shows a slice of the  $\K_x^{\rm we,p_2}$ kernel at about $100$~km above the photospere, and in (b) we show the same slice, but of the function obtained by  equation~(\ref{sep1}). In the bottom two panels depth slices of these two kernels are shown for comparison. The match between the kernels is quite good. We have also studied the point-to-point kernels in this way.  In general, the weighted azimuthally averaged kernels (the ones actually used in the inversion) separate `better' than the point-to-point ones, because much of the small-scale structure is averaged away. All of the other kernels for ridge filtering that we have studied separate this way to a very good approximation, giving us confidence in the type of inversion we choose to employ.

\subsection{Noise covariance matrix}

Travel times contain a significant amount of realization noise and a good
understanding of these noise properties  allows us to assign accurate errors
to the flow estimates.  It has been shown in previous time-distance inversions
 that it is important to take into account the
noise covariance matrix \cite{jensen2003,couvidat2005,couvidat2006}. \inlinecite{gizon2004} showed in detail how to compute model noise covariances of travel times.

We assume the solar oscillations are stationary and spatially homogeneous since we are restricting our study to the quiet Sun. We further assume that the noise between different ridge measurements $n$ and $n'$ is uncorrelated. This approximation is acceptable for the type of ridge filtering that is
used in this study. The
covariance matrix $\Lambda$ of the noise components $\cn_\trt$ from equation~(\ref{tt-k}) is  given by
\begin{equation}
\label{noise}
\Lambda^{\alpha\beta}_{n}(\br_i-\br_j;\Delta,\Delta')={\rm Cov}\left[\cn_\trt^{\alpha,n}(\br_i;\Delta),\cn_\trt^{\beta,n}(\br_j;\Delta')\right].
\end{equation}
This quantity has units of ${\rm s}^2$, and is computed according to equation~(28) in \inlinecite{gizon2004}. Example plots for the case when $n={\rm f}$ and for $\Delta=5$~Mm were shown in \inlinecite{jackiewicz2007a} using the same model. Similar features are seen for all mode ridges and distances considered here. Note that the covariance matrix elements of $\Lambda$ scale with the observation time $T$ as
$T^{-1}$.

%
%

\section{Basic strategy of the 2+1 dimensional subtractive optimally localized averages inversion}
\label{sec:basic}

The problem we wish to solve is to  estimate, for example, $\uu_x(\br,z)$ in
equation~(\ref{tt-k}), given the travel-time measurements, the sensitivity kernels, and
the noise covariance matrix.  We carry this out using a 2+1D SOLA inversion procedure. We formulate the problem in terms of the component $\uu_x$ only for notational simplicity, noting that the  procedure for finding all other flow components is completely equivalent. We first discuss the basic idea of this method, and then in the following two sections separately describe the 2D and 1D parts in more detail.

An OLA-type inversion for our purposes seeks a way to combine the sensitivity kernels to find a three-dimensional averaging kernel
that is, roughly, the shape of a ball, centered horizontally about  the origin $\br={\mathbf 0}$
and vertically about  the target depth $z_t$ located somewhere in the solar interior. The averaging kernel will be found from the 2+1D inversion and is defined by
\begin{equation}
\vv_x(\br;z_t)=h_r^2 h_z\sum_{j,z}{}^\tpod\bcK^{\uu_x}(\br_j-\br,z;z_t)\cdot\bbu(\br_j,z)+{\rm noise}\, ,
\label{avekerndef}
\end{equation}
where $\vv_x$ is an estimate of the real solar flow $u_x$ (as is the case in the rest of the paper) and the noise term will be specifically quantified below. To clarify the notation, sensitivity kernels are written as K and averaging kernels are written as $\cK$. Any superscript to the left of the averaging kernel denotes by which type of inversion it was computed, for example, ${}^{\oned}\cK$,  ${}^{\twod}\cK$, and ${}^{\tpod}\cK$. The superscripts to the right indicate for which  component of $\bbu$  and ridge $n$ the averaging kernel is computed. 

It is important to observe from equation~(\ref{avekerndef}) that the $y$ and  $z$ components of the averaging kernel should be zero so that the flow estimate $\vv_x$ is not contaminated by any cross talk from $\uu_y$ and $\uu_z$. As seen in the next section, the 2D inversion attempts to accomplish this by constraining the spatial integral of $\cK_y$ and $\cK_z$ to be zero. Ideally, one would want the $x$ component of the averaging kernel to be a delta function; however, noise and a finite set of travel times inhibit this. Nonetheless, if an acceptable averaging kernel is found, then the travel times can be properly averaged to give an estimate of the local flow $\vv_x\approx\uu_x$ in which we are interested.

Since the problem essentially separates into a 2D and a 1D problem because of ridge filtering,  the
three-dimensional averaging kernel ${}^\tpod\bcK^{\vv_x}$ from equation~(\ref{avekerndef}) is derived in two main steps. The first
step is to compute the 2D (horizontal) component of the averaging kernel, ${}^\twod\bcK(\br)$, such
that its $x$ components is highly peaked  about the point $\br={\mathbf 0}$, and the other components are zero. This typically involves trying to match the $x$ component  to a target function that is a 2D Gaussian function in horizontal coordinates $\br=(x,y)$. The inversion coefficients, or weights $w$, that accomplish this averaging of the sensitivity kernels, are then used to combine the travel times in such a way that  the estimated flow for any ridge measurements $n$ is
\begin{equation}
\vv^n_x(\br;\Delta)=\sum_{j,\alpha}w^{\alpha,n}(\br_j-\br;\Delta)\tau_n^\alpha(\br_j;\Delta). 
\label{vtt}
\end{equation}
The resulting flow map  $\vv^n_x$  is an average of the real flow $\uu_x$ over the depth that the
dominant modes of ridge $n$ probe. An intermediate step is to combine these maps  over all
distances $\Delta$ by an averaging procedure based on the correlated noise in the  measurements. A set of weightings $\gamma$ is computed, described in detail in  Section~\ref{sec:combine}, such that the distance-averaged flows are
given by
\begin{equation}
\vv^n_x(\br)=\sum_j \gamma_j^{n} \,\vv^n_x(\br;\Delta_j),
\label{ave-flows}
\end{equation}
where $\gamma_j$ is defined in equation~(\ref{comb_weights}) and the sum $j$ runs over all $\Delta$ used in the problem.

The second main step is to obtain localization of the  3D averaging kernel in the vertical
direction about $z_t$ by combining separate 2D ridge measurements. A 1D inversion in depth is thus performed which seeks a new set of inversion coefficients $c^n$.  These new
coefficients combine the 2D flow maps in equation~(\ref{ave-flows}) in such a way  that the final estimate
of the flow  around  $z_t$ is given by
\begin{equation}
\vv_x(\br;z_t)=\sum_n c^n(z_t)\vv^n_x(\br)\approx\uu_x(\br;z_t),
\label{vvn}
\end{equation}
where $\uu_x(\br;z_t)$ denotes the real flow at a particular depth $z_t$. The whole procedure is in principle carried out  to estimate each flow component $(\uu_x,\uu_y,\uu_z)$ at many different target depths $z_t$ to infer the vector flow $\bbu(\br,z)$ throughout a desired interior region. We now describe in more detail how  the 2D and 1D inversion weights are computed  in the following  two sections.

\section{2D horizontal inversion}

\begin{figure}
\centerline{\includegraphics[width=.8\textwidth]{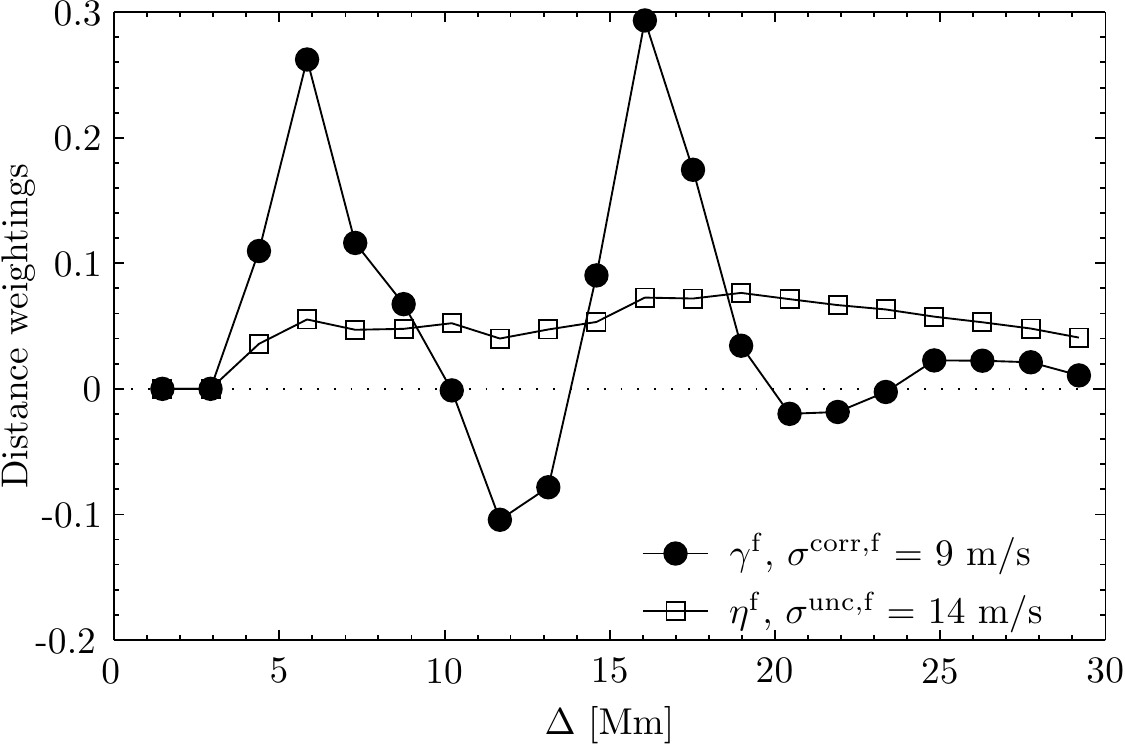}}
\caption{Example distance weightings  from a 2D f-mode inversion for $\uu_x$ using  a target resolution of $11$~Mm, as described in Section~\ref{sec:combine}. The observation time is $24$~h. The open squares are the weights at each distance obtained by assuming that the the estimated flows $\vv_x^{\rm f}(\br;\Delta)$ are uncorrelated for different $\Delta$ (eq.~[\ref{uncorr}]). The points with filled circles are the weights taking the correlations properly into account (eq.~[\ref{comb_weights}]). The noise $\sigma$ for each case indicated in the legend.}
\label{fig:dist_weights}
\end{figure}

\begin{figure}
\centerline{\includegraphics[width=\textwidth]{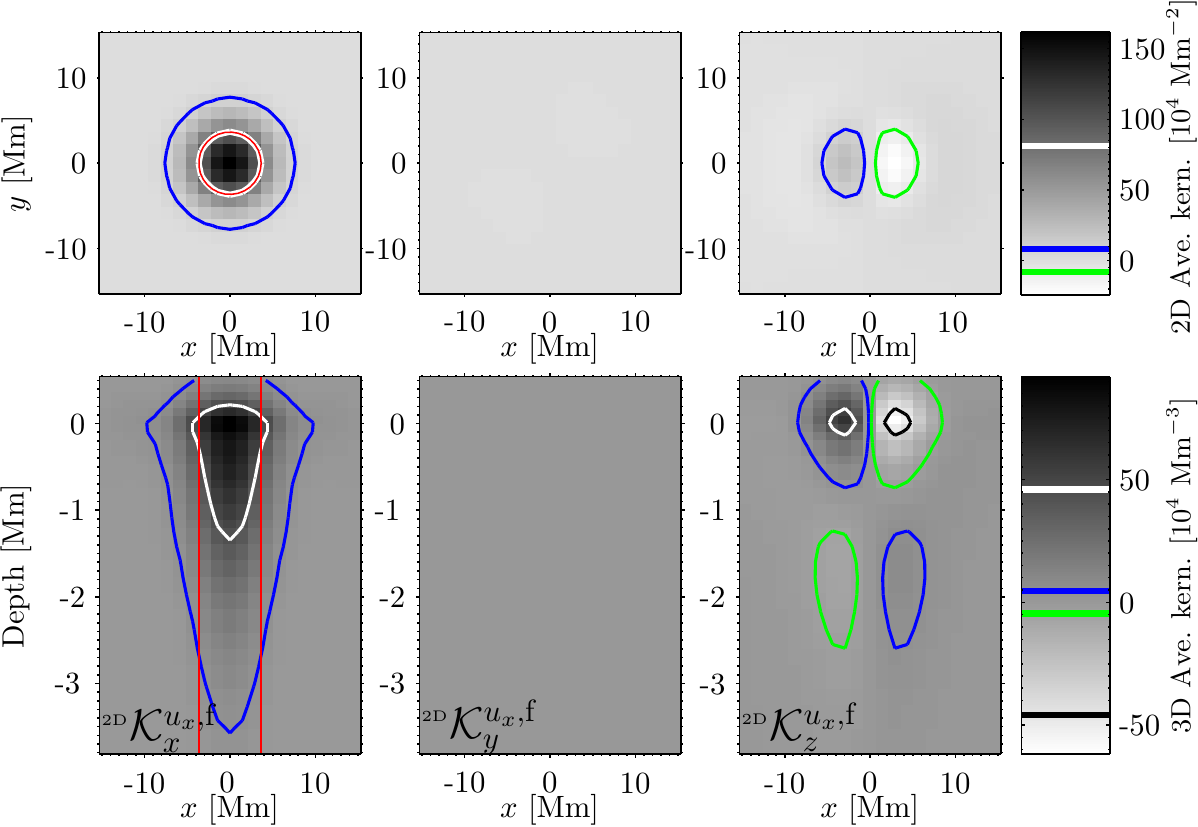}}
 \caption{Two-dimensional (top panels) and three-dimensional (bottom panels)
    averaging kernel for a 2D inversion for $\uu_x$ using only f modes. The
    panels in the top row show the integral over depth  of
    the corresponding  3D kernel below it. The bottom row shows depth slices of the three components of the 3D kernel along the $y=0$ line. The red line outlines the FWHM$=7.3$~Mm of the 2D Gaussian target function used in the inversion (see equation~[\ref{target}]). The overplotted  color contours, which are also marked on the colorbar for reference, denote the following: (white) is the half maximum of the kernel, (black) is the negative of the half maximum, (blue) and (green) denote $\pm 5$\% of the maximum value of the kernel, respectively. Estimates of the noise from this and all other inversions are given in Table~\ref{tab:weights}.}
 \label{fig:mode_avekerns_f_vx}
\end{figure}

\begin{figure}
\centerline{\includegraphics[width=\linewidth]{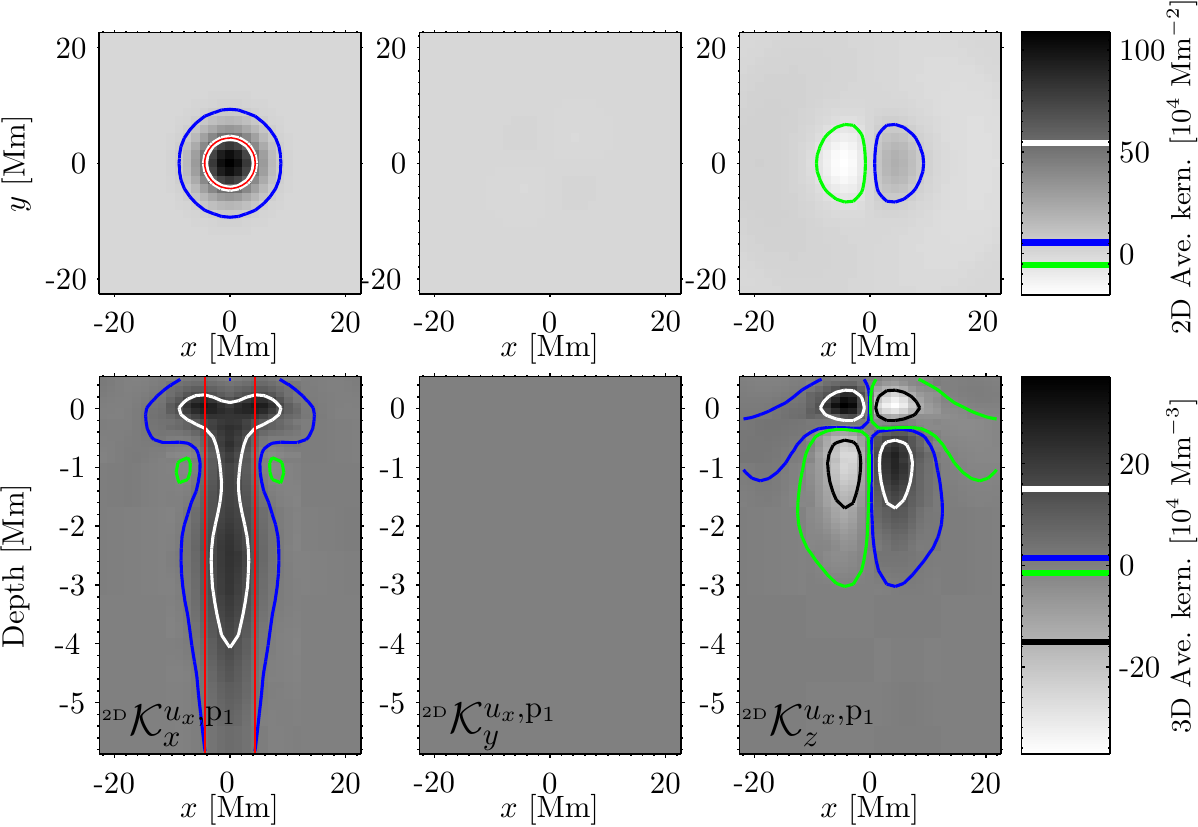}}
 \caption{Two-dimensional (top panels) and three-dimensional (bottom panels)
    averaging kernel for a 2D inversion for $\uu_x$ using only \pone\ modes.  The colors are as for Figure~\ref{fig:mode_avekerns_f_vx}.}
  \label{fig:mode_avekerns_p1_vx}
\end{figure}

Based on the separability of the sensitivity kernels discussed in Section~\ref{sec:forward}, the 2D inversion is formulated to solve equation~(\ref{tt-k}) for $\vv_x(\br)$ using the
2D depth-integrated kernel
\begin{equation}
\bK^{\alpha,n}(\br;\Delta)=h_z\sum_z \bK^{\alpha,n}(\br,z;\Delta),
\end{equation}
which we compute for all $\alpha$, $n$, and $\Delta$ available. 

We can define the averaging kernel that we wish to find from the 2D inversion, ${}^\twod\bcK^{\uu_x,n}$, by plugging equation~(\ref{tt-k}) into   equation~(\ref{vtt}) to obtain
\begin{equation}
\vv_x^n(\br;\Delta)=h_r^2h_z\!\!\sum_{i,z}{}^\twod\bcK^{\uu_x,n}(\br_i-\br,z;\Delta)\cdot\bbu(\br_i,z)+\sum_{j,\alpha}w^{\alpha,n}(\br_j-\br;\Delta)\cn_{\rm tt}^{\alpha,n}(\br_j;\Delta),
\label{2davekern}
\end{equation}
where
\begin{equation}
{}^\twod\bcK^{\uu_x,n}(\br,z;\Delta)=\sum_{j,\alpha}w^{\alpha,n}(\br_j;\Delta)\bK^{\alpha,n}(\br-\br_j,z;\Delta).
\label{3davekern}
\end{equation}
This shows explicitly that the averaging kernel gives only an estimate $\vv_x^n$ of the flow that is some average of the real flows over some depth.

The details for
obtaining the inversion  weights $w$ for a 2D SOLA inversion for flows were
presented  in \inlinecite{jackiewicz2007a}, however, in that work only one mode ridge, one distance $\Delta$, and two components of the sensitivity kernels (${\rm K}_x$ and ${\rm K}_y$) were utilized.  The generalization to our present case is straightforward. To summarize this procedure,  we first prescribe a target function ${}^\twod\bcT^{\uu_x}$ that we wish the averaging kernel to resemble. It is a vector-valued function, chosen such that the $x$ component  is typically a 2D Gaussian in $\br$ with dispersion $\sigma$, and the other components are zero:
\begin{equation}
{}^\twod\bcT^{\uu_x}(\br) = \left(\frac{{\rm e}^{-r^2/2\sigma^2}}{2\pi\sigma^2},0,0\right),
\label{target}
\end{equation}
where  $r=||\br||$ is the 2D vector norm.  The horizontal integral of the target function is normalized to one. The full-width at half-maximum (${\rm FWHM}=2\sigma\sqrt{2\ln 2}$) of the target function is a measure of the resolution of the inversion if the averaging kernel matches it well. For the sake of completeness, in an inversion for the $j$th component of $\bbu$, and denoting the gaussian function in equation~(\ref{target}) as $G(r)$, the $i$th component of the target function is ${}^\twod\bcT_i^{\uu_j}=G(r)\hat{\mathbf e}_i\delta_{ij}$, where $\hat{\mathbf e}_i$ is the unit vector in the $i$th direction and $\delta$ is the Kronecker delta function.

Two quantities, one which measures the mismatch between the averaging
kernel and target function, the other which measures the noise propagation,
are computed.  Let the noise in the inversion for $\uu_x$ for ridge $n$ be denoted by $\cn^{\uu_x,n}$; it is specifically defined in Section~\ref{sec:combine}. A minimization (with respect to the inversion weights) is carried out according to
\begin{equation}
\min_w\sum_i ||{}^\twod\bcK^{\uu_x,n}(\br_i)-{}^\twod\bcT^{\uu_x}(\br_i)||^2 + \beta\left(\cn^{\uu_x,n}\right)^2,
\label{min2d}
\end{equation}
where $\beta$ is some regularization parameter that we choose  typically to be quite small \cite{jackiewicz2007a}.   A large matrix is then regularized and inverted for each value of the trade-off parameter, which results in a unique set of weights at each point in this parameter space. We choose  an `optimal' set of weights $w$ from examining the
trade-off curve (L curve)  as discussed in \inlinecite{jackiewicz2007a}, such that the averaging kernel matches closely the target function. The 2D averaging kernel is constructed by convolution of the sensitivity kernels with the weights according to equation~(\ref{3davekern}).

 Since the inversion  in this example is carried out for $\uu_x^n$, it is important that
${}^\twod\cK_y^{\uu_x,n}$ and ${}^\twod\cK_z^{\uu_x,n}$ be as close to zero as possible, to
minimize the `cross talk' among  all of the components. This is achieved in
practice by constraining the total spatial integrals of the $y$ and $z$
components to be zero, although in practice there is usually some structure
present even with this constraint. We are in the process of exploring other  effective constraints. When a well-localized averaging kernel is found for each ridge $n$, the weights are then suitable
to be used to average the travel times to give an estimate of the flow
$\vv_x^n(\br)$ using equation~(\ref{vtt}).

\subsection{Combining all of the distances $\Delta$}
\label{sec:combine}

Throughout this inversion procedure, it is necessary  to combine the quantities 
we obtain for different  annulus radii $\Delta$, such as  the estimated flow maps
$\vv_x^n(\br;\Delta)$. This is done by weighting each distance by appropriate weights.  One simple way of achieving this, typically used in helioseismology, is to  assume that the noise in each measurement is independent and uncorrelated. Then the standard deviation in the estimated flow maps is used to determine the contribution of the errors at each distance. We denote the standard deviation of a set of flows $\vv_x^n$ for each distance $\Delta_j$ and mode ridge $n$ as $\sigma_j^{\sigmaunc,n}$. The `unc' superscript emphasizes the assumption of uncorrelated data. Finding the  minimum variance of this set then gives a weighting factor, $\eta_j^n$, according to
\begin{equation}
\eta_j^n=\frac{\left(1/\sigma_j^{\sigmaunc,n}\right)^2}{\sum_{j=1}^{20}\left(1/\sigma_j^{\sigmaunc,n}\right)^2},
\label{uncorr}
\end{equation}
where the sum in the denominator runs over all 20 distances used in this problem. We carry out a 2D inversion as described above for $\uu_x$ using f modes  in  a region of quiet Sun (the same region used in Section~\ref{sec:results}).  The weights obtained from the estimated flows using equation~(\ref{uncorr}) are plotted in Figure~\ref{fig:dist_weights} as open squares. For very small distances where the noise is very high, the weights are zero. From the total variance we obtain a noise estimation given by
\begin{equation}
\sigma^{\sigmaunc,{\rm f}}=\left[\sum_j \left(1/\sigma_j^{\sigmaunc,{\rm f}}\right)^2\right]^{-1/2},
\end{equation}
which for this particular example is  $14$~m\,s$^{-1}$.

However, we know that the values of the flows at
different $\Delta$ are correlated quite strongly due to noise \cite{gizon2004}, and so we choose to average them in a way that takes these correlations into account. A covariance matrix $C_n$ of the noise in the individual flow measurements at distances $\Delta$ and $\Delta'$ of
ridge $n$ is
computed using the 2D inversion weights as
\begin{equation}
C_n(\Delta,\Delta') = \sum_{i,j,\alpha,\beta}w^{\alpha,n}(\br_i;\Delta)\Lambda_{n}^{\alpha\beta}(\br_i-\br_j;\Delta,\Delta')w^{\beta,n}(\br_j;\Delta'),
\label{ccov}
\end{equation}
where  $\Lambda$ is the covariance matrix of the noise in the travel times
(equation~[\ref{noise}]), and $w$ are the 2D inversion weights.  Note that the
matrix $C$ has units of ${\rm length}^2{\rm time}^{-2}$. The final measurement
of any general  quantity $q$ for each mode ridge
is then obtained by averaging the $20$ distances  we use according to (for example, see \opencite{schmelling1995}):
\begin{equation}
  q^n(\br) = \sum_{j=1}^{20} \gamma_j^n q^n(\br;\Delta_j),
\label{vcomb}
\end{equation}
where the set of weightings $\gamma_j^n$ is given by
\begin{equation}
\gamma_j^n = \frac{\left(\sum_{i=1}^{20}C_n^{-1}(\Delta_i,\Delta_j)\right)}{\left(1/\sigma^{\sigmacorr,n}\right)^2},
\label{comb_weights}
\end{equation}
and the variance $(\sigma^{\sigmacorr,n})^2$ for the correlated case is
\begin{equation}
\left(\sigma^{\sigmacorr,n}\right)^2 =\left(\sum_{ij}\left(C^{-1}_n(\Delta_i,\Delta_j)\right)\right)^{-1}.
\label{noisecorr}
\end{equation}
We can now identify $\cn^{\uu_x,n}\equiv\sigma^{\sigmacorr,n}$, the noise in a measurement of $\vv_x^n$, introduced in equation~(\ref{min2d}). A set of weights for the f-mode case $\gamma_j^{\rm f}$ obtained this way are plotted as the filled circles in  Figure~\ref{fig:dist_weights} to compare with the uncorrelated case. What is interesting to note is that for some distances a negative contribution is needed to average the data properly, which is never the case for the uncorrelated data. The estimated noise is also lower than in the uncorrelated case. Equation~(\ref{vcomb}) is quite general, and has been used to obtain the distance-averaged flow maps presented in Section~\ref{sec:results}, as well as all of the averaging kernels shown in this paper. Quantities written without the distance argument $\Delta$ have been averaged this way.



\subsection{Averaging kernels from the 2D inversion}
\label{sec:2davekerns}

\begin{figure}
\centerline{\includegraphics[width=\linewidth]{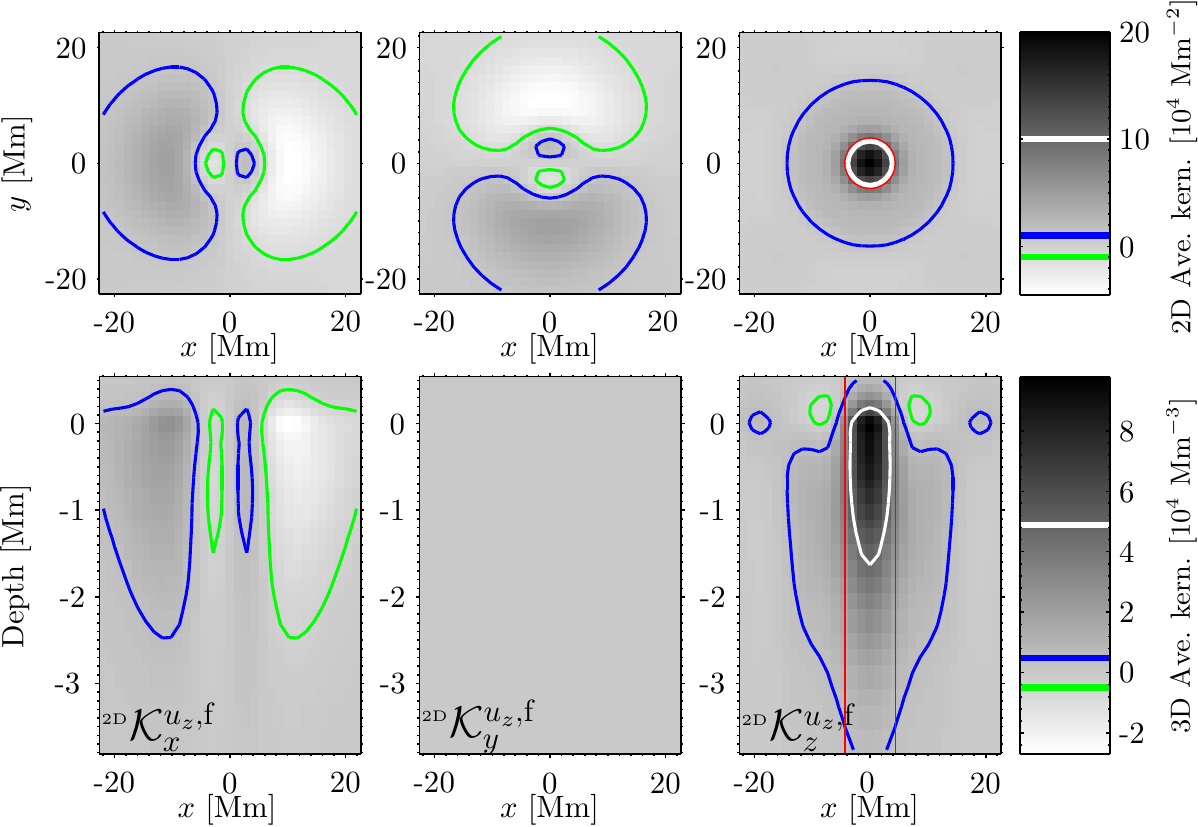}}
\caption{Two-dimensional (top panels) and depth slices of three-dimensional (bottom panels)
    averaging kernel for a 2D inversion for $\uu_z$ using only f modes.  The colors are as for Figure~\ref{fig:mode_avekerns_f_vx}.}
 \label{fig:mode_avekerns_f_vz}
\end{figure}

\begin{figure}
\centerline{\includegraphics[width=\linewidth]{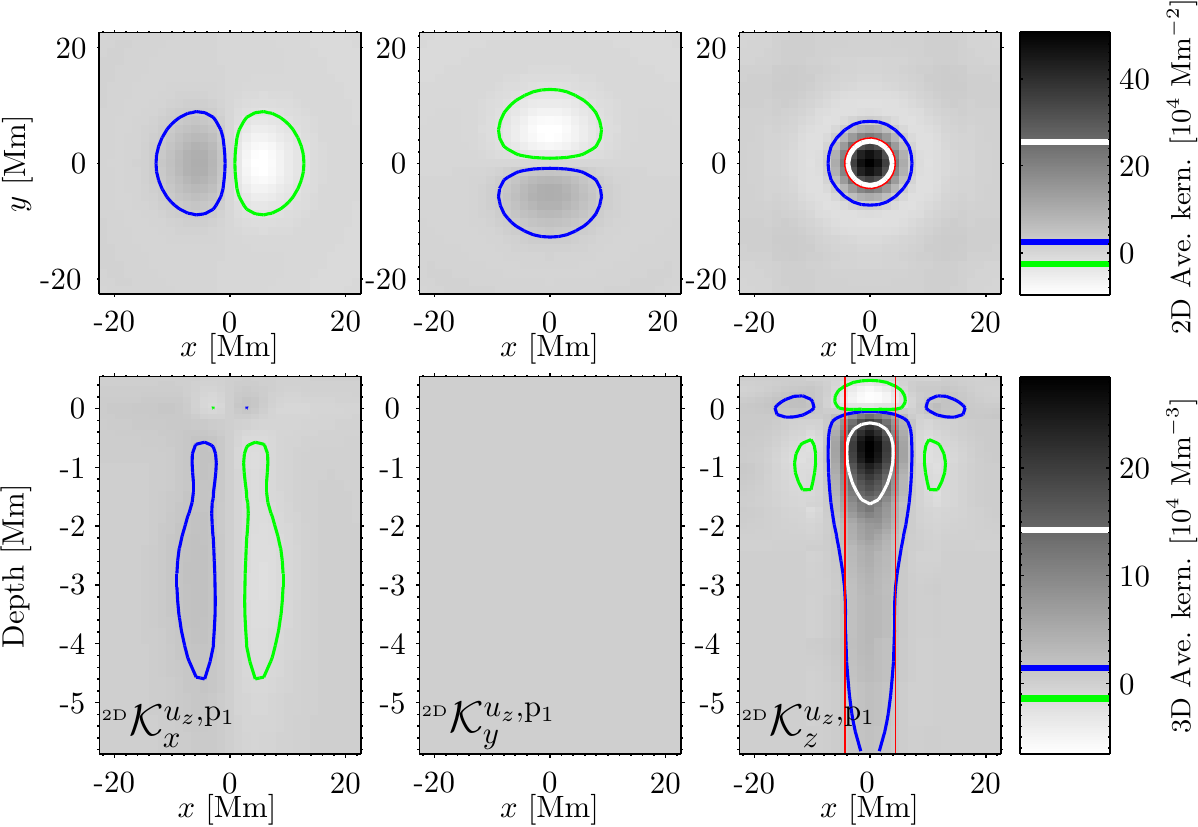}}
 \caption{Two-dimensional (top panels) and depth slices of three-dimensional (bottom panels)
    averaging kernel for a 2D inversion for $\uu_z$ using only \pone\ modes.  The colors are as for Figure~\ref{fig:mode_avekerns_f_vx}.}
  \label{fig:mode_avekerns_p1_vz}
\end{figure}

Example averaging kernels after combining all distances for a 2D inversion 
 are shown in
Figures~\ref{fig:mode_avekerns_f_vx}-\ref{fig:mode_avekerns_p1_vz}. In all
figures, the top row shows the 2D averaging kernel obtained by integrating over
depth the adjacent 3D averaging kernel in the bottom row. The bottom panels show depth
slices along $y=0$. The inherent noise from the inversion corresponding to each figure is given in Table~\ref{tab:weights}.

Figures~\ref{fig:mode_avekerns_f_vx} and \ref{fig:mode_avekerns_p1_vx} show averaging kernels for a 2D inversion for $\uu_x$ for the f and \pone\ mode ridges, respectively, while Figures~\ref{fig:mode_avekerns_f_vz} and
\ref{fig:mode_avekerns_p1_vz} are for an inversion for $\uu_z$. The absence of
any dominant  cross talk is evident
in the top panels of all figures, which are the 2D averaging kernels
that come straight out of the 2D inversion. There is a completely  negligible $y$
component in all cases for the inversions for $\uu_x$. The cross talk is slightly more pronounced  for the $x$ and $y$ components of the kernels in the inversions for $\uu_z$, but is confined to the very near-surface region, typically above the depths which are significant for our inversion results. In
general, these averaging kernels are quite good; the only structure from the `off diagonal' terms are of the order of about $5$\% of the diagonal terms. It is important to study averaging kernels such as these to have an idea of what the inversion is actually accomplishing. Similar plots have been examined for
all of the other mode ridges available and they exhibit similar features.

\subsection{Minimum variance}
\label{sec:minvar}

\begin{figure}
\centerline{\includegraphics[width=\linewidth]{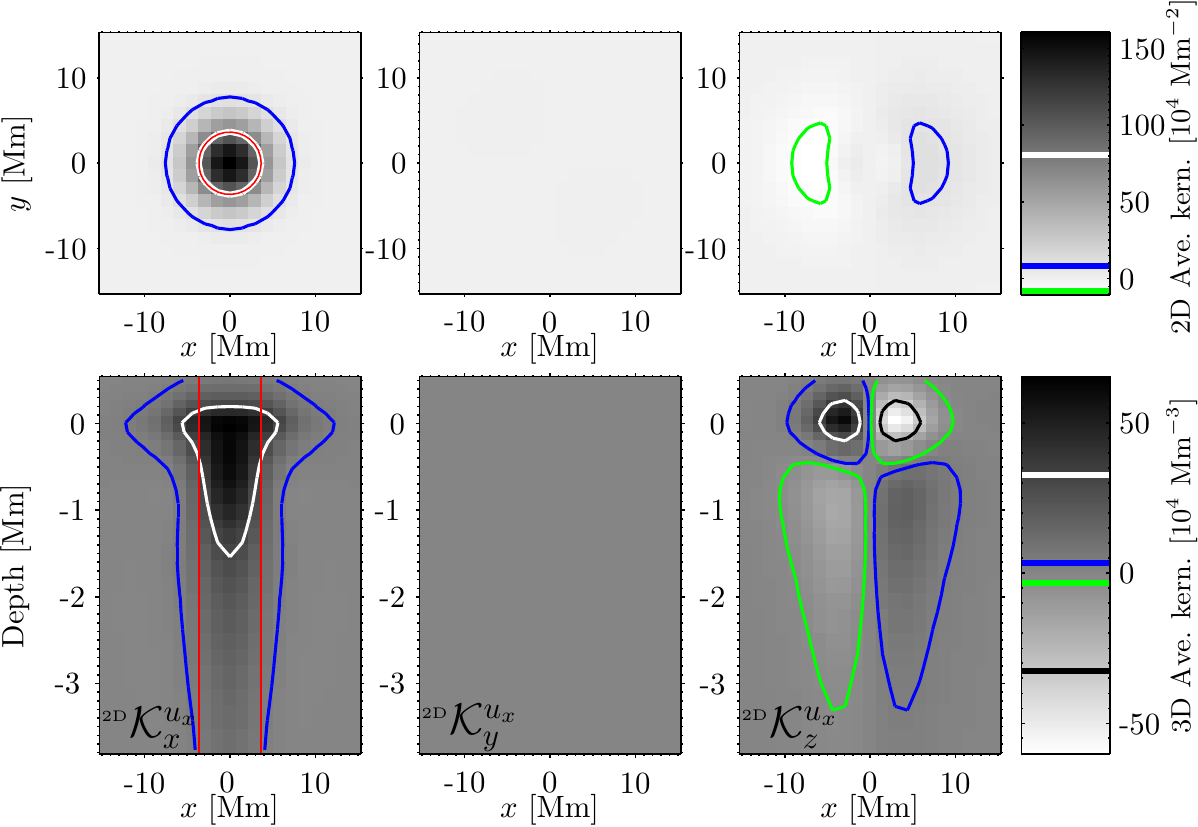}}
 \caption{Three-dimensional minimum variance averaging kernel for a 2D
    inversion for $\uu_x$. The top panels are the 2D averaging kernels after integrating the 3D kernel over depth, and the
    bottom panels are slices at  $y=0$ through the 3D kernel. The method to obtain this kernel is discussed in Section~\ref{sec:minvar}. The resolution of the inversion is $7.3$~Mm. The colors are as for Figure~\ref{fig:mode_avekerns_f_vx}.}
  \label{fig:3davekerns_minvar}
\end{figure}

The averaging kernels  from the 2D inversion in
Figures~\ref{fig:mode_avekerns_f_vx}-\ref{fig:mode_avekerns_p1_vz} are computed for  each separate mode ridge
$n$. One possible way to combine them over all ridges is to use a simple minimum
variance treatment of the noise in the estimated flow component $\vv_x^n$, as described in Section~\ref{sec:combine} for the case of distance averaging. Because of ridge filtering,  we consider the noise between mode ridges $n$ and $n'$ to be uncorrelated. Thus, any quantity $q^n$ that depends on the set of mode ridges can be averaged according to
\begin{equation}
q^{\uu_x}=\sum_n d^n q^{\uu_x,n},
\end{equation}
where the weights $d^n$ are given by
\begin{equation}
d^n=\frac{\left(1/\sigma^{\sigmacorr,n}\right)^2}{\sum_n\left(1/\sigma^{\sigmacorr,n}\right)^2},
\label{minvarweights}
\end{equation}
and $\sigma^{\sigmacorr,n}$, the correlated noise estimated from the $\vv_x^n$  measurements, is defined in equation~(\ref{noisecorr}).  In
Figure~\ref{fig:3davekerns_minvar} we show the three components of an averaging kernel $\bcK^{\uu_x}$
obtained by combining five kernels $\bcK^{\uu_x,n}$ from the minimum variance
in $\vv_x^n$ using the weights given in equation~(\ref{minvarweights}). The minimum-variance weights for each ridge and noise for this figure are given in Table~\ref{tab:weights}. The noise level
is low  and the kernel is  well localized horizontally, however,  we clearly  have no control over the depth at which one wishes to have sensitivity.  

In conclusion, the 2D inversion produces averaging functions (such as those shown in
Figures~\ref{fig:mode_avekerns_f_vx}-\ref{fig:3davekerns_minvar}) that  are
`optimally' localized in the horizontal direction, but not in depth. This depth localization is
accomplished by  a 1D inversion, to which we now turn.

\section{1D SOLA depth inversion}
\label{sec:1d}

\begin{figure}
\centerline{\includegraphics[width=\linewidth]{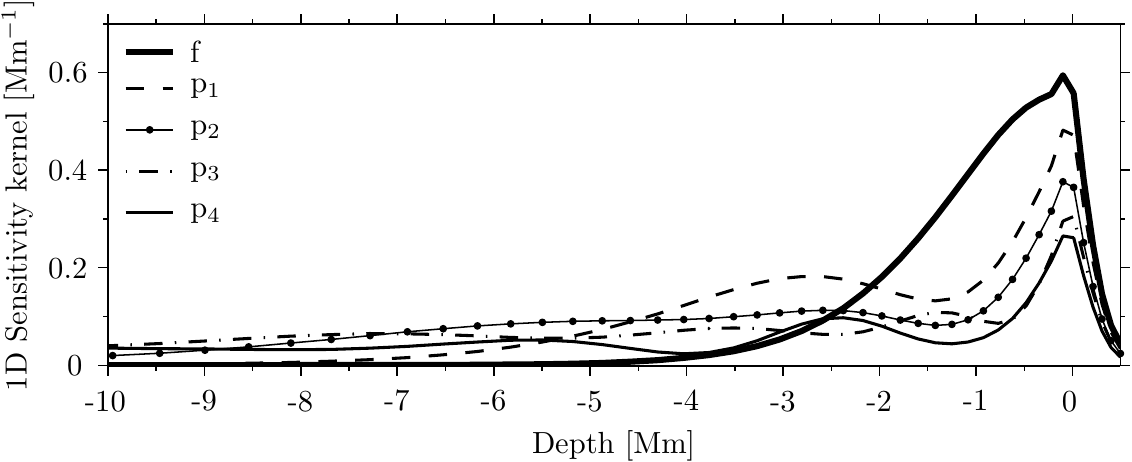}}
 \caption{One-dimensional sensitivity kernels $^\oned\K_x^{\uu_x,n}(z)$ as a function of depth for each ridge $n$ (indicated in the legend) for an inversion for $\uu_x$. These are obtained  according to equation~(\ref{1dsenskern}).}
\label{fig:oned_kerns}
\end{figure}

\begin{figure}
\centerline{\includegraphics[width=.8\linewidth]{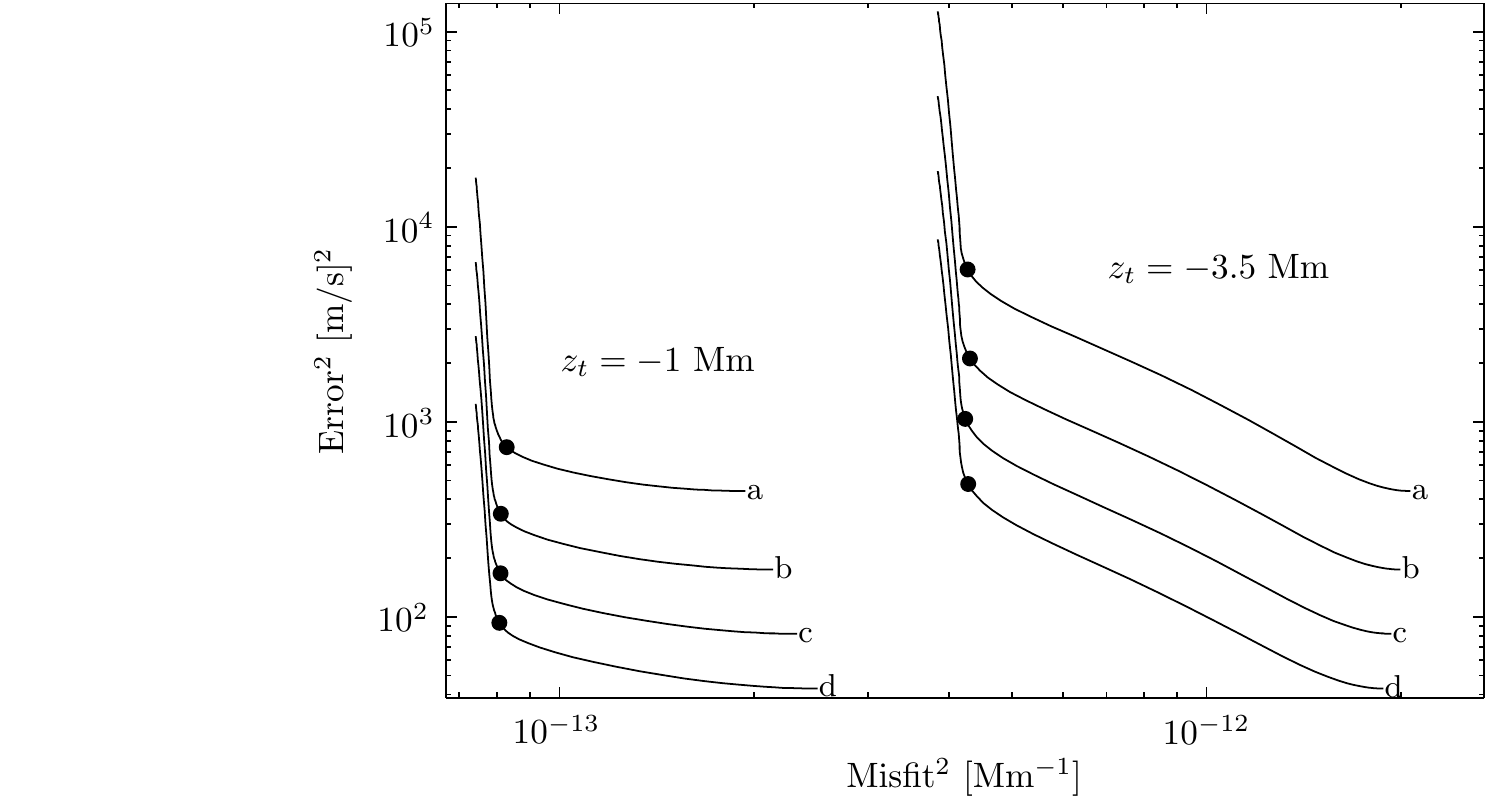}}
\caption{Example trade-off curves computed in the 1D inversion for two target
               depths. The individual curves in each set of depths are obtained by varying the horizontal target
               function  width. For the $z_t=-1$~Mm set, curves a, b, c, and d correspond to $5.8$, $7.3$, $8.7$, and $10.2$~Mm, respectively. For the $z_t=-3.5$~Mm, a, b, c, and d corresond to $7.3$, $8.7$, $10.2$, and $11.6$~Mm, respectively.  In this scheme, sets of inversion weights would be
               chosen at the points given by the circles.}
\label{fig:lcurves}
\end{figure}

Up to this point, the 2D inversion has provided averaging kernels which average the real solar flows to give estimates  of the flows for each particular ridge, according to
equation~(\ref{2davekern}).
If we assume that the real flows vary slowly in $\br$ over the horizontal extent of the averaging kernel, we can perform the summation in  equation~(\ref{2davekern}) over $\br_i$ to obtain
\begin{equation}
\vv_x^n(\br)\approx h_z\sum_z{}^{\oned}\K_x^{\uu_x,n}(z)\uu_x(z)+\sum_{j,\alpha}w^{\alpha,n}(\br_j-\br)\cn_{\rm tt}^{\alpha,n}(\br_j),
\label{v1d}
\end{equation}
where
\begin{equation}
{}^\oned\K_x^{\uu_x,n}(z) := h_r^2\sum_i{}^{\twod}\cK^{\uu_x,n}_x(\br_i,z),
\label{1dsenskern}
\end{equation}
is the  1D sensitivity kernel for an inversion for $\uu_x$ and mode ridge $n$. Recall that the horizontal integrals of ${}^\twod\cK_y$ and  ${}^\twod\cK_z$ are zero due to the constraint imposed in the 2D inversion; therefore, only the $x$ component of the quantities remains in the right hand sides of equation~(\ref{v1d}) and (\ref{1dsenskern}).

The five available 1D sensitivity kernels are shown in
Figure~\ref{fig:oned_kerns}. The 1D SOLA inversion seeks  inversion coefficients
$c^n(z_t)$ that combine each ridge measurement about target depth $z_t$, so that 
the final estimate of the flow using equation~(\ref{v1d}) is
\begin{eqnarray}
\nonumber
\vv_x(\br;z_t)&=&\sum_n c^n(z_t)\vv_x^n(\br)\\\nonumber
&=& h_z\sum_z{}^\oned\cK_x^{\uu_x}(z;z_t)\uu_x(\br,z)+\sum_{j,\alpha,n} c^n(z_t)w^{\alpha,n}(\br_j-\br)\cn_{\rm tt}^{\alpha,n}(\br_j),
\end{eqnarray}
where
\begin{equation}
{}^\oned\cK_x^{\uu_x}(z;z_t) = \sum_n c^n(z_t){}^\oned\K_x^{\uu_x,n}(z)
\label{1davekerns}
\end{equation}
is the one-dimensional averaging kernel peaked about $z_t$. The 1D coefficients $c^n$ are obtained in an analogous way to  the 2D case.  A target function is chosen which is typically a 1D Gaussian in depth, centered about $z_t$:
\begin{equation}
{}^\oned\cT_x^{\uu_x}(z;z_t)=\frac{{\rm e}^{-(z-z_t)^2/2\sigma^2}}{\sigma\sqrt{2\pi}}.
\end{equation}
A misfit quantity is constructed  which measures the mismatch between the target and averaging functions:
\begin{equation}
{\rm misfit}^2=h_z\sum_z\left[{}^\oned\cK_x^{\uu_x}(z;z_t)-{}^\oned\cT_x^{\uu_x}(z;z_t)\right]^2.
\end{equation}
In addition, a quantity which quantifies the error (noise)  is included:
\begin{equation}
{\rm error}^2=\sum_n \left(c^n\cn^{\uu_x,n}\right)^2.
\end{equation}
A regularization parameter $\mu$ is introduced, and a minimization procedure with respect to the weights $c^n$ is carried out as
\begin{equation}
\min_{c}\left[{\rm misfit}^2 + \mu\,{\rm error}^2\right].
\end{equation} 
Computing the minimization  results in a system of linear equations, which is solved by inverting a small matrix  to obtain the coefficients for each
value of the regularization parameter. Finally, we choose weights roughly in
the `elbow'  of the trade-off curve upon visual inspection. Several example
trade-off curves from this procedure are shown in Figure~\ref{fig:lcurves} for
two target depths $z_t$ and different target widths.

\begin{figure}
\centerline{\includegraphics[width=.8\linewidth]{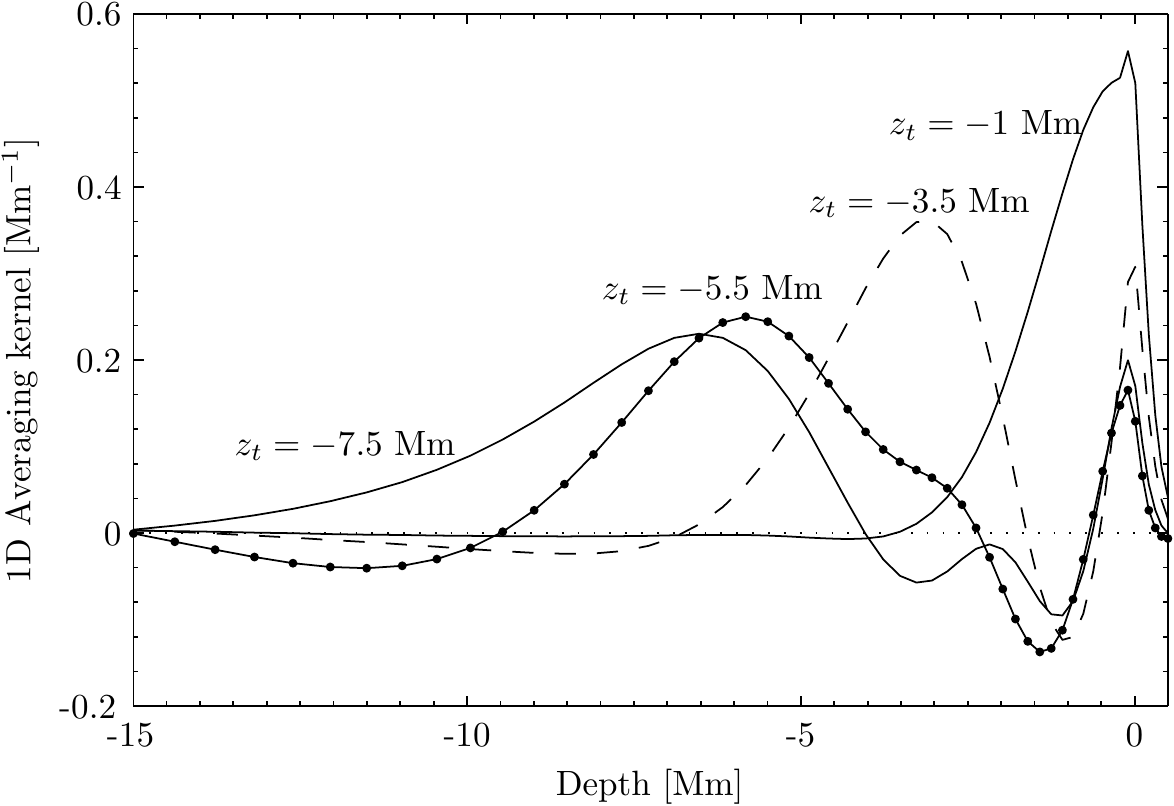}}
\caption{1D averaging kernels $^\oned\cK^{\uu_x}(z;z_t)$ for an inversion for $\uu_x$ for different target depths. These are obtained according to equation~(\ref{1davekerns}).}
\label{fig:1davekerns}
\end{figure}

In Figure~\ref{fig:1davekerns} we provide examples of  1D averaging kernels for an inversion for $\uu_x$ and for different target depths. Due to
a limited mode set, there is a limited number of depths that can be targeted
properly. There is also an obvious limit on the maximum depth with which we can probe
with these modes. In addition, as with other  helioseismology inversions, a
surface component is  present (see, e.g., \opencite{basu1999}).

\section{3D averaging kernels from the 2+1D inversion}
\label{sec:3davekerns}

\begin{figure}
\centerline{\includegraphics[width=\linewidth]{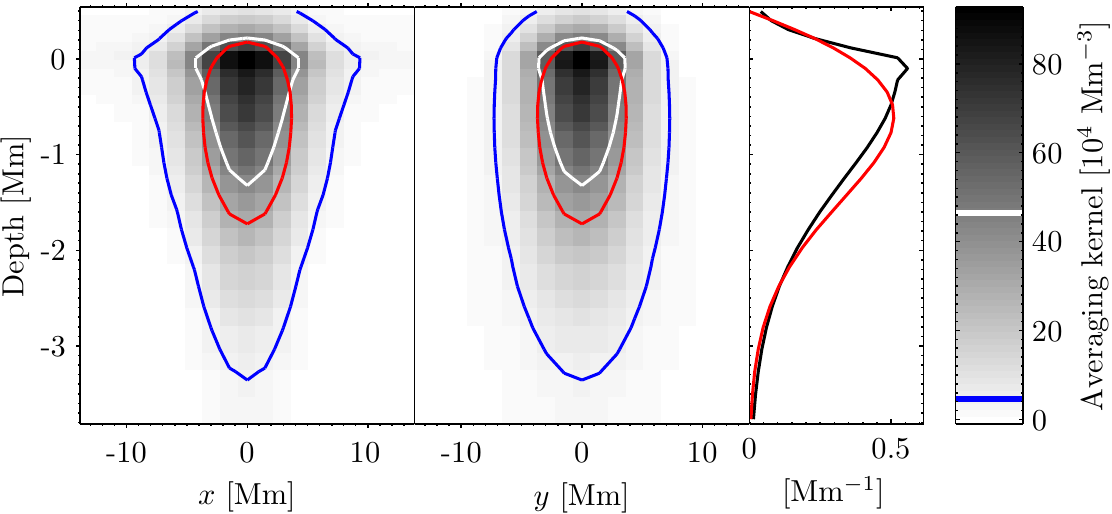}}
 \caption{Vertical slice through a 3D  averaging kernel ${}^\tpod\cK_x^{\uu_x}$ after performing a 2+1D
   inversion for $\uu_x$ for a target depth of about $-1$~Mm below the solar
   surface. The left (middle) column is a slice along the $y=0$ ($x=0$)
   line. The red contours show the half maximum value of the 3D target function, white
   contours show $50$\% of  the maximum value of the averaging kernel, black contours show $-50$\% of the maximum value of the averaging kernel, and the
   blue (green) contours denote $\pm 5$\% of the maximum. In the rightmost panel, the red line is the 1D target function, and
   the black solid line is the 1D averaging kernel.  The 1D weights $c^n$ and
   the noise estimate are given in Table~\ref{tab:weights}. }
\label{fig:avek_1}
\end{figure}

\begin{figure}
\centerline{\includegraphics[width=\linewidth]{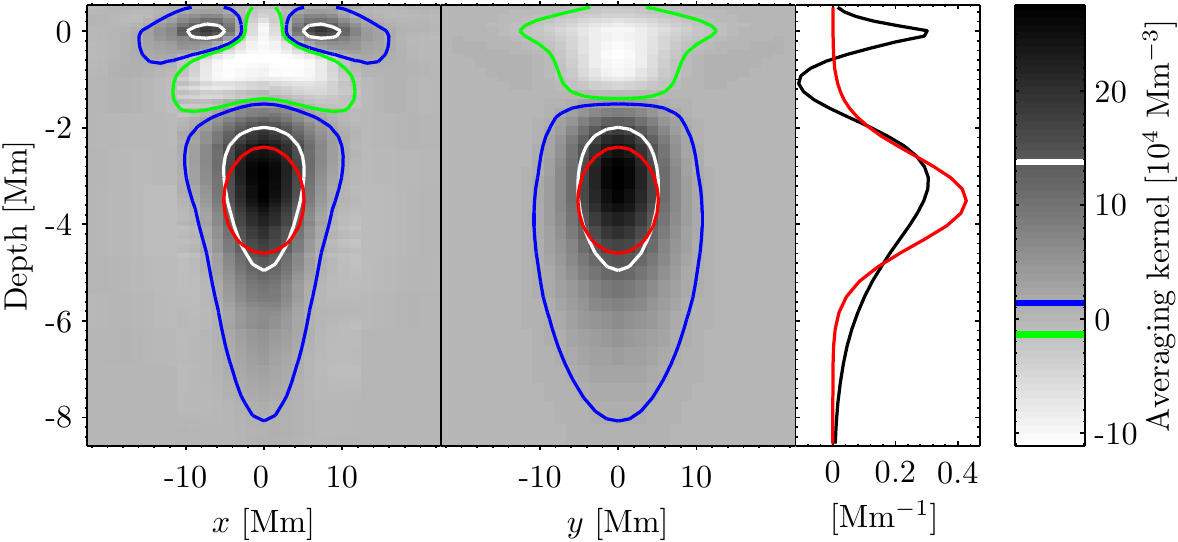}}
 \caption{Vertical slices through an averaging kernel after performing a 2+1D
   inversion for $\uu_x$ for target   depth $z_t\approx-3.5$~Mm. The right panel shows the 1D averaging kernel and target function. The colors are the same as for Figure~\ref{fig:avek_1}.}
\label{fig:avek_3.5}
\end{figure}

\begin{figure}
\centerline{\includegraphics[width=\linewidth]{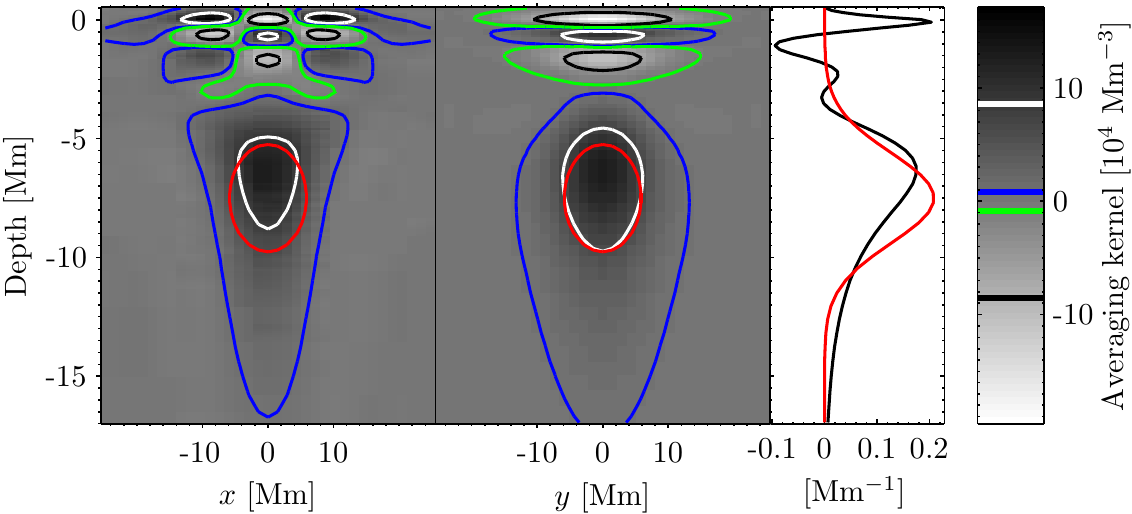}}
 \caption{Vertical slices through an averaging kernel after performing a 2+1D
   inversion for $\uu_x$ for  target depth $z_t\approx-7.5$~Mm. The right panel shows the 1D averaging kernel and target function. The colors are the same as for Figure~\ref{fig:avek_1}.}
\label{fig:avek_7.5}
\end{figure}

We denote the final 3D averaging kernel produced from the 2+1D inversion for $\uu_x$ as ${}^\tpod\bcK^{\uu_x}$.  It has been defined in equation~(\ref{avekerndef}), and can be constructed from both sets of inversion coefficients ($w$, $c$) in  terms of the original sensitivity kernels using equations~(\ref{vvn}) and (\ref{2davekern}):
\begin{eqnarray}
{}^\tpod\bcK^{\uu_x}(\br,z;z_t)&:=&\sum_n c^n(z_t){}^\twod\bcK^{\uu_x,n}(\br,z)\\
&:=& \sum_{ij\alpha n}c^n(z_t)\gamma^n_i w^{\alpha,n}(\br_j;\Delta_i)\bK^{\alpha,n}(\br-\br_j,z;\Delta_i),
\end{eqnarray}
where we emphasize that the weights $w$ and $c$ are obtained from a specific inversion for $\uu_x$. Recall that the weights $\gamma$ are used to average the quantities over distance $\Delta$.

We now check to see if the final averaging kernels are as well localized as can be expected from the mode set used, which also justifies separating the problem into  2D and 1D parts. Performing the full 2+1D inversion for $\uu_x$ for different target depths produces example 3D averaging kernels such as those
shown in  Figures~\ref{fig:avek_1}, \ref{fig:avek_3.5}, and
\ref{fig:avek_7.5}. Plotted in the left and center panels in each case are depth slices along the $y=0$ and $x=0$ lines of the $x$ component of the kernel, ${}^\tpod\cK_x^{\uu_x}$. The contour of the half-maximum value  of the 3D target function is
overplotted on the depth slices in red. The white contours show the half-maximum value of the  3D averaging kernel. Note that in principle, in a noiseless inversion with a large set of available modes, the white contours would match perfectly the red contours. The blue and green contour lines denote  $\pm 5$\% of the maximum value. Also provided in the rightmost panel
in each figure for comparison are the 1D target and averaging functions. For the shallowest target depth, instead of a simple 1D gaussian we use a target function that goes to zero at $z=0$. The 1D inversion
coefficients $c^n$ used to construct the kernels in  Figures~\ref{fig:avek_1} -- \ref{fig:avek_7.5} are provided in Table~\ref{tab:weights}. The shallowest
depth we can reach with these modes is about $-1$~Mm, the deepest about
$-8$~Mm. Of course, the noise begins to increase quickly with depth.

\begin{table}
\caption{Values of the relevant quantities for some of the figures in this
  paper. Listed is the figure number, the type of inversion used to produce
  the figure, the 1D inversion weights $c^n$ (or minimum variance weights $d^n$), and the estimated noise associated with each measurement ($1\sigma$ values are given). Note that the 1D inversion weights do not apply for the 2D inversion. }
\label{tab:weights}
\begin{tabular*}{\textwidth}{@{\extracolsep{\fill}}lcrrrrrr}
\hline
Figure~\# & Inversion type & $c^{\rm f}$ & $c^{\rm p_1}$ & $c^{\rm p_2}$ & $c^{\rm p_3}$ & $c^{\rm p_4}$ & noise \\\hline
\ref{fig:mode_avekerns_f_vx}  & 2D  & - & - & - & - & - & 16~m\,s$^{-1}$\\
\ref{fig:mode_avekerns_p1_vx} & 2D & - & - & - & - & - & 26~m\,s$^{-1}$\\
\ref{fig:mode_avekerns_f_vz}  & 2D & - & - & - & - & - & 17~m\,s$^{-1}$\\
\ref{fig:mode_avekerns_p1_vz}  & 2D & - & - & - & - & - & 24~m\,s$^{-1}$\\
\ref{fig:3davekerns_minvar} & 2D~(min.~var.) & 0.65 & 0.24 & 0.086 & 0.014 & 0.002 & 21~m\,s$^{-1}$ \\
\ref{fig:avek_1} & 2+1D & 1.02 & -0.04 & -0.01 & 8.2e-04 & 6.4e-04 & 27~m\,s$^{-1}$\\
\ref{fig:avek_3.5} & 2+1D & -1.2 & 2.2 & -9.7e-04 & -0.04 &  -2.7e-04 & 40~m\,s$^{-1}$ \\
\ref{fig:avek_7.5} & 2+1D & -0.17 &  -1.7 &  2.3 &  0.5 &  0.05 & 57~m\,s$^{-1}$ \\
\ref{fig:mode_comp2}~(a) & 2+1D & 0 & 1 & 0 & 0 & 0  & 19~m\,s$^{-1}$\\
\ref{fig:mode_comp2}~(b) & 2+1D & 0.33 & - &  0.47 & 0.17 & 0.05 & 17~m\,s$^{-1}$\\
\ref{fig:3depths} (top) & 2+1D & 1.08 &   -0.13 & 0.004 &  0.01 & 0.005 & 10~m\,s$^{-1}$\\
\ref{fig:3depths} (middle) & 2+1D & -0.11 &   1.01 & 0.03 &  0.04 & 0.01 & 12~m\,s$^{-1}$\\
\ref{fig:3depths} (bottom) & 2+1D & -0.92 &   1.26 & 0.63 &  0.02 & 0.01 & 22~m\,s$^{-1}$\\
\hline
\end{tabular*}
\end{table}

%
%
\section{Results with MDI data for quiet-Sun flows}
\label{sec:results}
In the rest of the paper we provide example flow maps in the quiet Sun from the 2D and 2+1D
inversion procedure discussed above. It is our main intention to demonstrate
that the results obtained are sensible and consistent with what might be
expected given the spatial resolution, observation time, and level of estimated noise. We note also that in order to study local flows, we typically remove a mean, large-scale, time-averaged flow from each retrieved map.

\begin{figure}
  \centerline{\includegraphics[width=\linewidth]{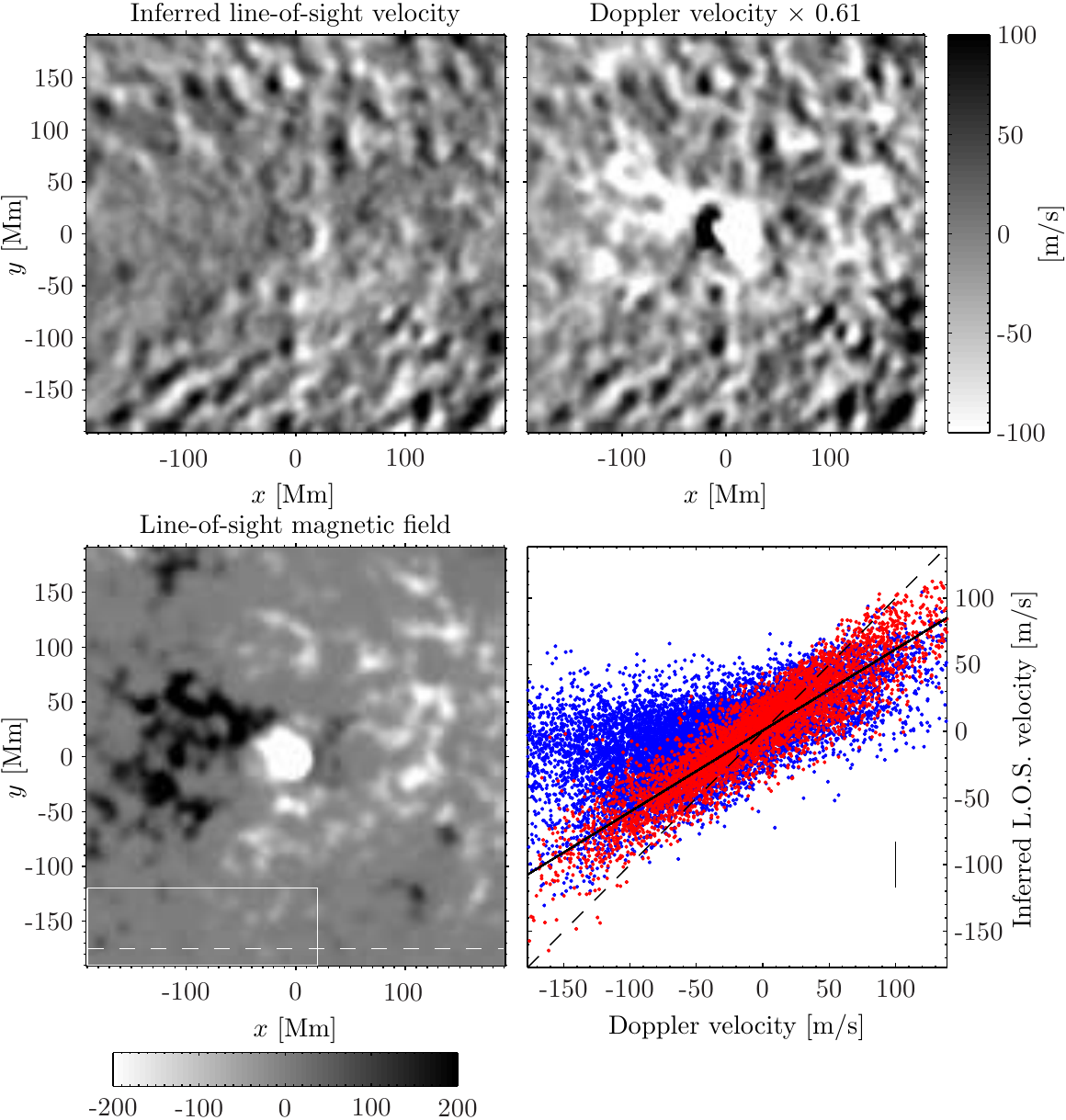}}
  \vspace{.2cm}
  \caption{Test of the 2D inversion. The top panels show a  comparison of the inverted flows with the direct Doppler data for a
    map centered at $30$ degrees west of disk center. The
  top left panel is the inferred vector flow map after  projection onto the line-of-sight vector at
  each pixel for an inversion using f modes and $24$~h of data. The resolution
  is about $7$~Mm. The Doppler
  map is 
  an average over $1$~day of MDI full-disk data, smoothed  with the 2D averaging
  kernel from the inversion  and  multiplied by the slope $0.61$ of the best fit line through the scatterplot (panel below)  to allow for direct comparison with the inverted
  map. The lower left panel is the 1-day averaged (truncated) MDI
  magnetogram with a sunspot visible in the middle. The units are in gauss. The white box outlines the quiet-Sun region analyzed for all of the plots in the rest of the paper. The white dashed line shows the location of the slice for the flows in Figure~\ref{fig:depthslice}. The lower right panel
  shows a scatterplot of the velocity maps, where red (blue) dots represent pixels of
  less (more) than $|10|$~G. The correlation is 0.9 for the non-magnetic data. The dashed line is $y=x$, and the solid line a
  fit to the non-magnetic data (slope $0.61$). The small vertical line represents
  the $\pm 1\sigma$ ($\sigma=17$~m\,s$^{-1}$) noise in the flow estimation.}
  \label{fig:flows_los}
\end{figure}

The first simple test we perform is to  compare the inferred flows from the 2D inversion to the direct MDI Doppler data. To accomplish this, the three components of the inferred 2D vector  flows are projected onto the line-of-sight vector at each pixel. We
use $24$~h of data from the seventh day (Jan. 26 2002) of the nine-day data set
available. There is  a sunspot in the center of the map which is centered
vertically about the
equator and at 30 degrees towards the western solar limb for this day.
Figure~\ref{fig:flows_los} shows the comparison with the Doppler map from a 2D
inversion using only f modes. The correlation between the inferred flows and the Doppler flows is $0.9$  for pixels with less than $|10|$~G of magnetic field. Also provided is a magnetogram to show the locations of
a large sunspot, the surrounding plage, and the region of quiet Sun that is studied in all of the plots in the rest of the paper.  The scatterplot of the two velocity maps shows that the magnetic field introduces an anomalous 
second component to the velocity field, reinforcing our reason to restrict all further analyses to the quiet Sun.  That the inferred line-of-sight velocity is smaller than the
Doppler velocity in not surprising. It is most likely attributable
\cite{gizon2000,braun2004} to the average depth over which the flows are
measured. We cannot make a purely `surface' measurement with the available
mode set.

\subsection{Tests of the 2D inversion}
\label{sec:tests2d}

\begin{figure}
  \centerline{\includegraphics[width=.8\linewidth]{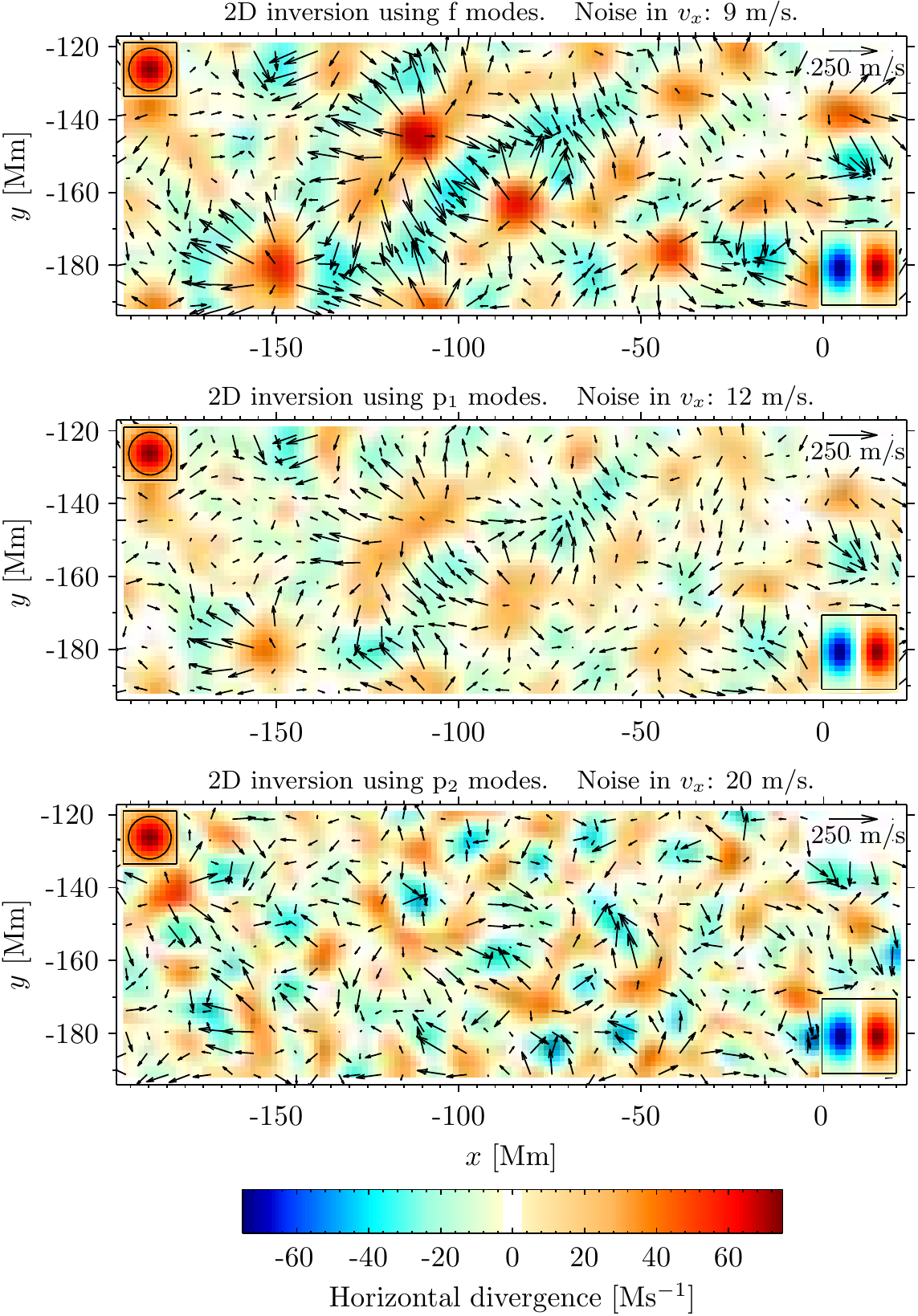}}
  \vspace{.5cm}
  \caption{Horizontal flows and divergence from a 2D inversion using $24$~hr of data for the f, \pone, and \ptwo\ ridges. In each panel, the arrows denote the horizontal flows (obtained from inverting for $\vv_x$ and $\vv_y$) for $24$~h of
    data, and the color scale is the horizontal flow divergence, obtained from a separate inversion as described in the text at the end of Section~\ref{sec:tests2d}. The $x$ component of the 2D averaging kernel from the flow inversion is plotted in the box in the upper left of each panel and the FWHM, outlined
    by the circle, is $11.6$~Mm. The $x$ component of the 2D averaging kernel for the horizontal divergence is given by the quantity in the lower right box (see Eq.~[\ref{divh}]). Note the strong supergranular flows in the f-mode map which gradually weaken as the modes probe deeper layers of the convection zone. The correlation of the f-mode map with the \pone-mode map is 0.88, and that between the f-mode and \ptwo-mode maps is 0.35. The  noise in $\vv_x$ is given for each panel. The noise in the horizontal divergence inversion is $10$, $12$, and $13$~Ms$^{-1}$ for the top, middle, and bottom panels, respectively.}
  \label{fig:3modes}
\end{figure}

\begin{figure}
  \centerline{\includegraphics[width=\linewidth]{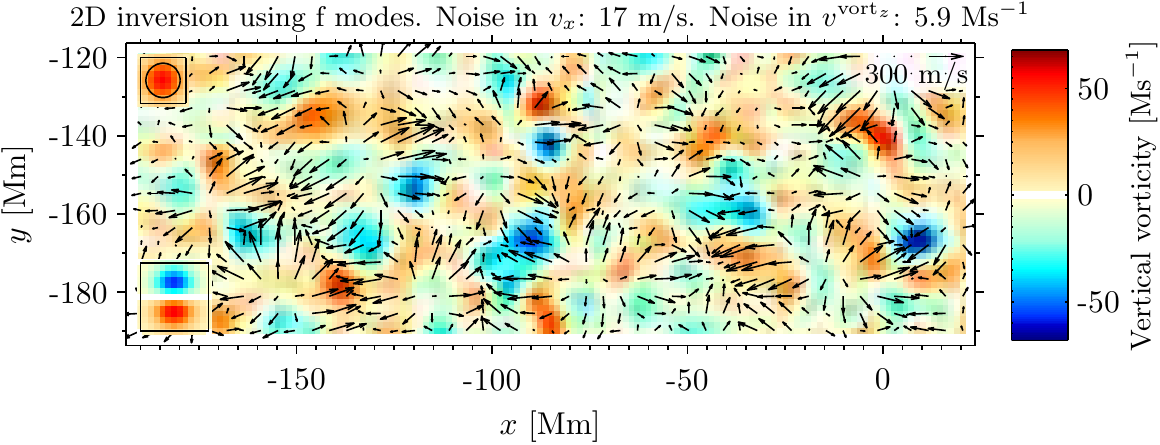}}
  \caption{2D inversions for the vertical vorticity and horizontal flows in the quiet Sun. The arrows denote the  horizontal flows and the color scale is the vertical vorticity obtained from a separate inversion using $24$~h or data. The $x$ component of the averaging kernel from the inversion for the horizontal flows is given in the upper left box, and the $x$ component of the averaging kernel for the vorticity inversion, which matches the target function from equation~(\ref{vortz}), is shown in the lower left box.}
  \label{fig:curl}
\end{figure}

\begin{figure} 
  \centerline{\includegraphics[width=.9\linewidth]{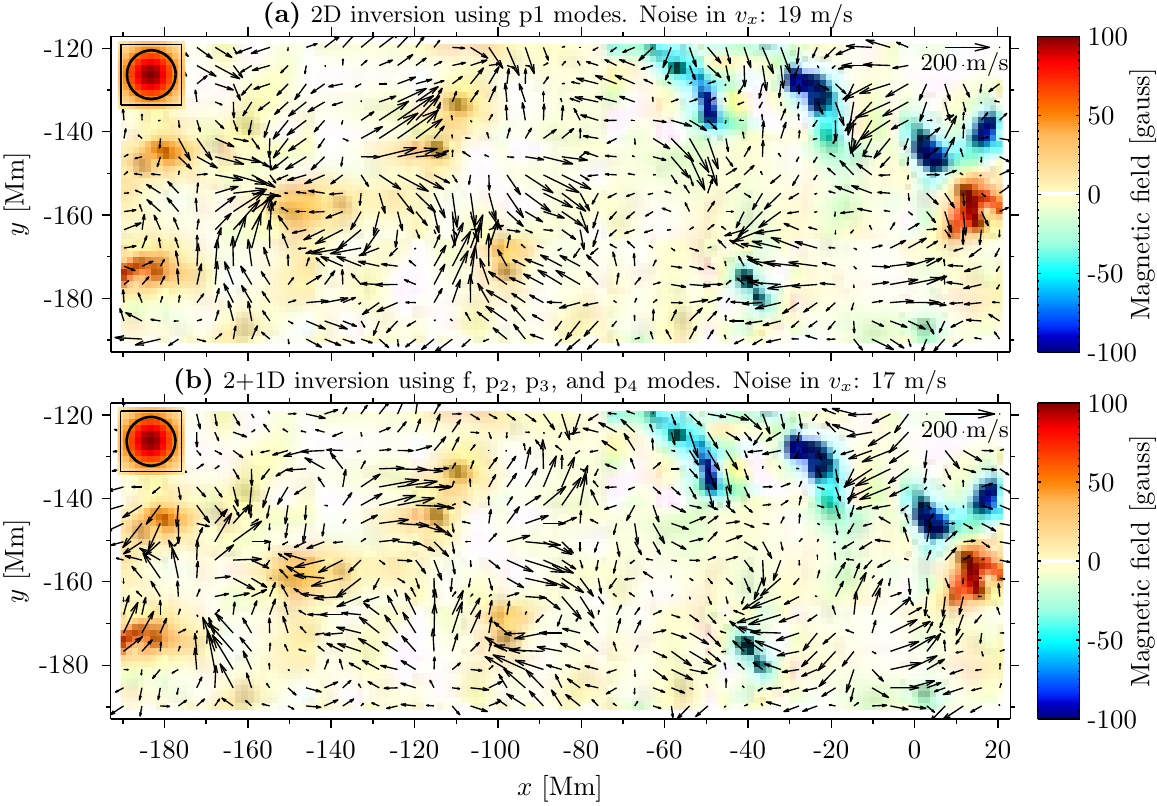}}
  \centerline{\includegraphics[width=.9\linewidth]{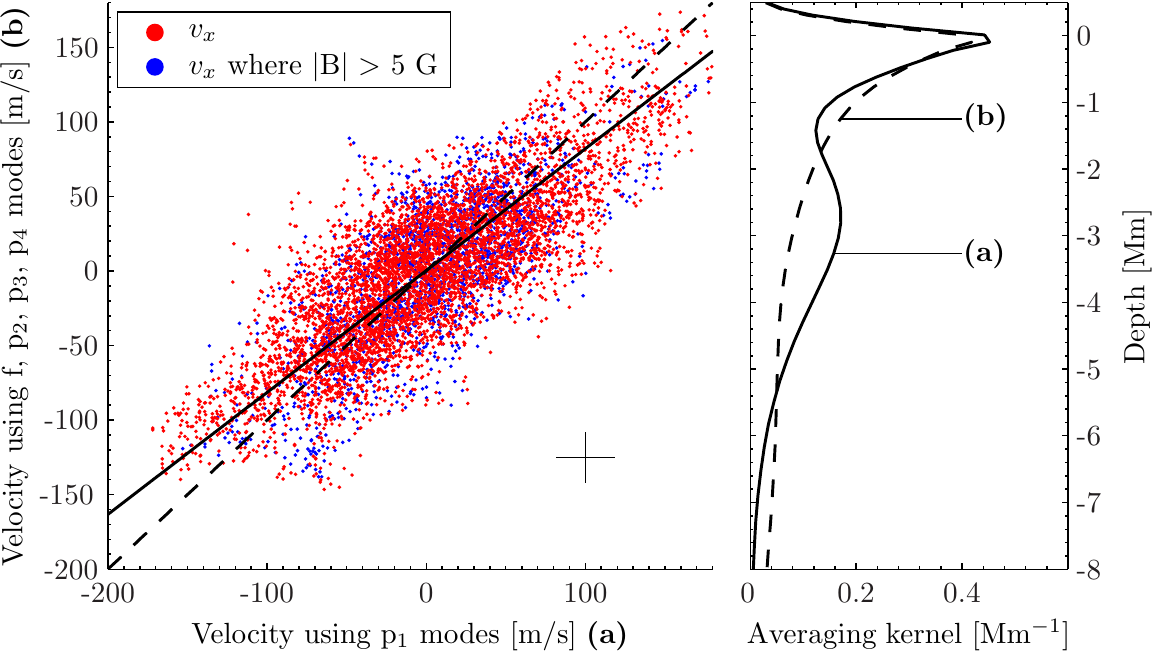}} 
  \caption{Test of the full 2+1D inversion. Panel~(a) shows the
    flows from a 2D inverison using only the \pone-mode ridge travel
    times. Panel~(b) shows flows from a 2+1D inversion using travel times from all
    other mode ridges (no \pone\ modes) by attempting to target the same
    averaging kernel. The arrows denote the horizontal flows (obtained from inverting for $\vv_x$ and $\vv_y$) for $24$~h of
    data, and the color scale is the truncated magnetic field from MDI. The $x$ component of the 2D averaging kernel is plotted in the box in the upper left, and the FWHM$=11$~Mm is outlined
    by the circle. The  scatterplot shows the flows from panel~(a) vs. (b), where the red (blue) dots are the values in the maps for which the
    magnetic field of the pixel is less (greater) than $|5|$~G. The dashed line is the $y=x$ line and the the solid line is the best
    fit through the non-magnetic data (slope$=0.83$), taking account of the errors on each
    axis. The
    $1\sigma$ error
    bars for each measurement are denoted by the cross in the lower right of
    the plot. The correlation in the scatter is $0.82$. The bottom right panel
    shows the 1D averaging kernel (solid line) for panel~(a) and the 1D averaging
    kernel (dashed line) for panel~(b). The 1D inversion weights for these two inversions are listed in Table~\ref{tab:weights}.}
\label{fig:mode_comp2}
\end{figure}

\begin{figure}
\centerline{\includegraphics[width=.9\linewidth]{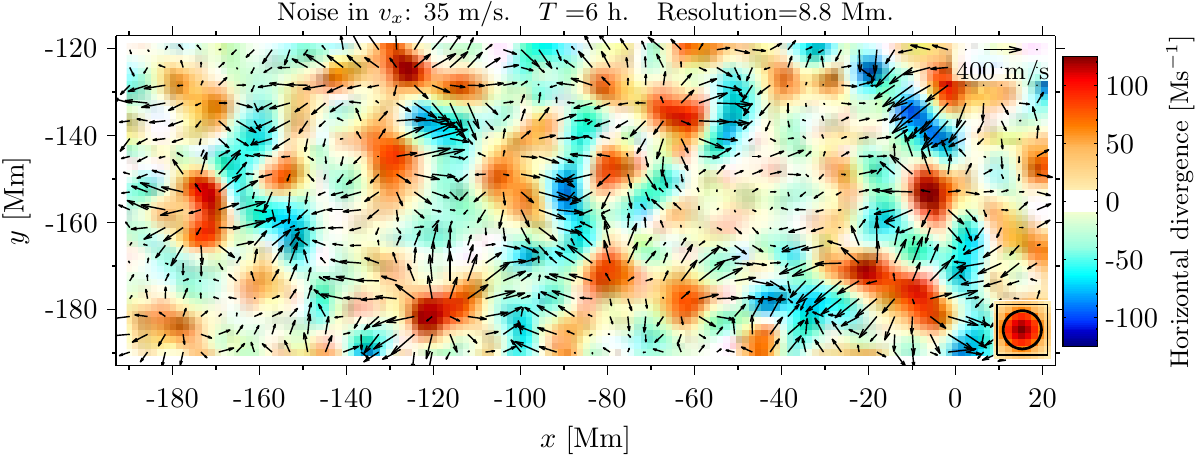}}
\vspace{-.5cm}
\centerline{\includegraphics[width=.9\linewidth]{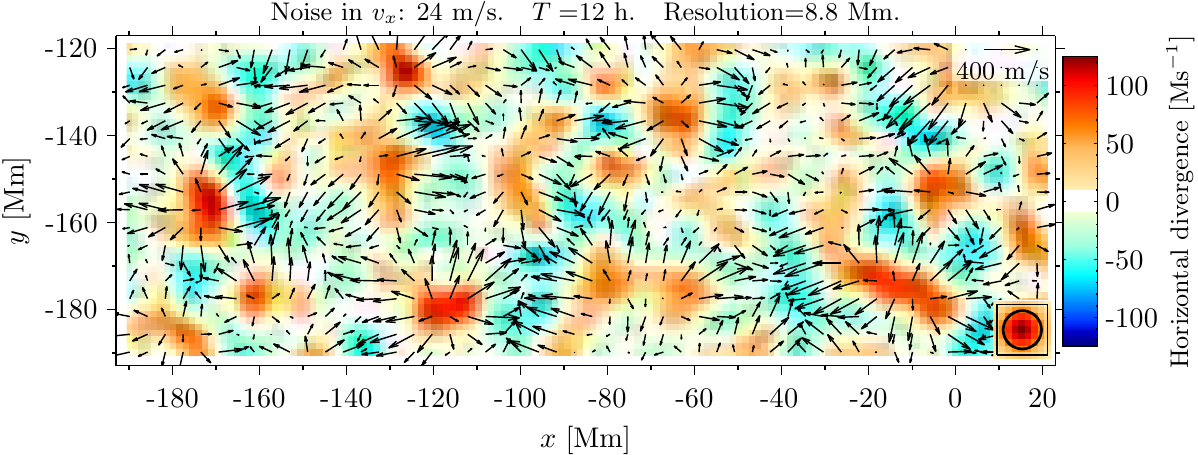}}
\vspace{-.5cm}
\centerline{\includegraphics[width=.9\linewidth]{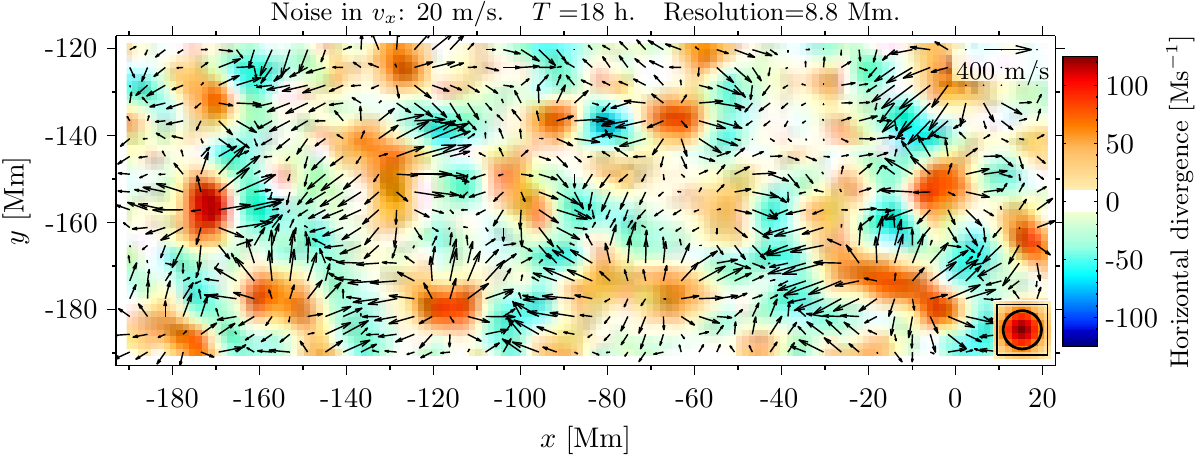}}
\vspace{-.5cm}
\centerline{\includegraphics[width=.9\linewidth]{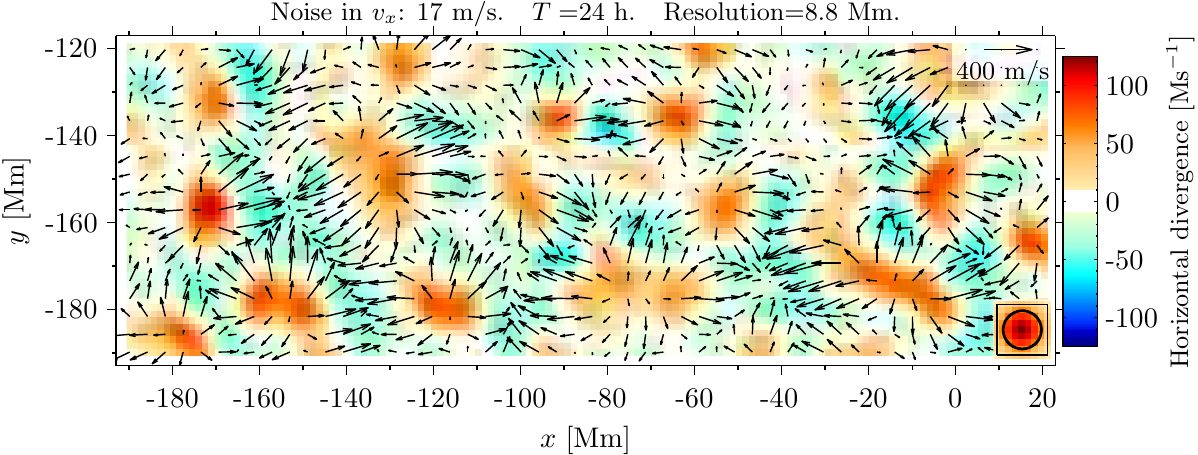}}
\caption{Comparison of flows for different observation times $T$ obtained from 2+1D inversions at a depth of $1$~Mm beneath the surface. The arrows are the inferred horizontal flows and the color scale is the horizontal divergence computed numerically. The total observation time increases from top to bottom in $6$~h intervals, but the resolution remains fixed. The correlation of the $6$~h map with the $24$~h map is 0.78, while the correlation of the $12$~h ($18$~h) map with the $1$~day map is 0.9 (0.97). Note that the divergence scale is the same for each panel.}
\label{fig:timeave}
\end{figure}

A further example of the 2D inversion is shown in Figure~\ref{fig:3modes}. Here we show horizontal flow maps inferred from inverting individual ridge travel times for f, \pone, and \ptwo\ and for $24$~h. The colorscale of these images is the horizontal divergence, obtained from a separate inversion as discussed below. In the f-mode map we see signatures of supergranulation with strong horizontal divergence in the centers of the supergranules and flow convergence at the cell boundaries. The outflow is generally in the $200-350$~m\,s$^{-1}$ range. The horizontal flows obtained from \pone\ and \ptwo\ travel times are weaker, as is the supergranulation signature. Maps for the \pthree\ and \pfour\ ridges have also been studied, but the noise begins to increase quickly with these modes at this resolution.

Obtaining the horizontal flow divergence from a direct inversion such as  shown  by the color scale in the plots in Figure~\ref{fig:3modes} is conveniently done in an OLA inversion such as this one. All that is needed is to use a different 2D target function such that the quantity that is inverted for is simply the horizontal divergence.

The function we use (which would replace the function defined in Eq.~[\ref{target}] in the 2D inversion) is
\begin{equation}
\bcT^{\rm div_h}(\br)=\left(\frac{x}{2\pi\sigma^4}e^{-r^2/2\sigma^2}, \frac{y}{2\pi\sigma^4}e^{-r^2/2\sigma^2}, 0 \right),
\label{divh}
\end{equation}
where ${\rm div_h}$ is a superscript label that denotes the horizontal divergence of the flow, $\bnabla_{\rm h}\cdot\bbv_{\rm h}$. Another useful quantity we have studied is the vertical component of the flow vorticity, $\hat{z}\cdot\bnabla_{\rm h}\times\bbv_{\rm h}$. We label the associated target function with the superscript ${\rm vort}_z$, which can be shown to be
\begin{equation}
\bcT^{{\rm vort}_z}(\br)=\left(\frac{-y}{2\pi\sigma^4}e^{-r^2/2\sigma^2}, \frac{x}{2\pi\sigma^4}e^{-r^2/2\sigma^2}, 0 \right).
\label{vortz}
\end{equation}
An example map of the vertical vorticity obtained directly from a 2D inversion is shown in Figure~\ref{fig:curl}.
We have studied plots of the divergence and vorticity at different resolutions and checked that the inverted quantities using these two types of targets and the direct numerical computation of these quantities using inferred horizontal vector flows agree reasonably well. The advantage of computing them directly from an inversion is that we are able to determine the noise and  spatial resolution properly.

\subsection{Tests of the 2+1D inversion}
\label{sec:tests2+1}

\begin{figure}
  \centerline{\includegraphics[width=.9\linewidth]{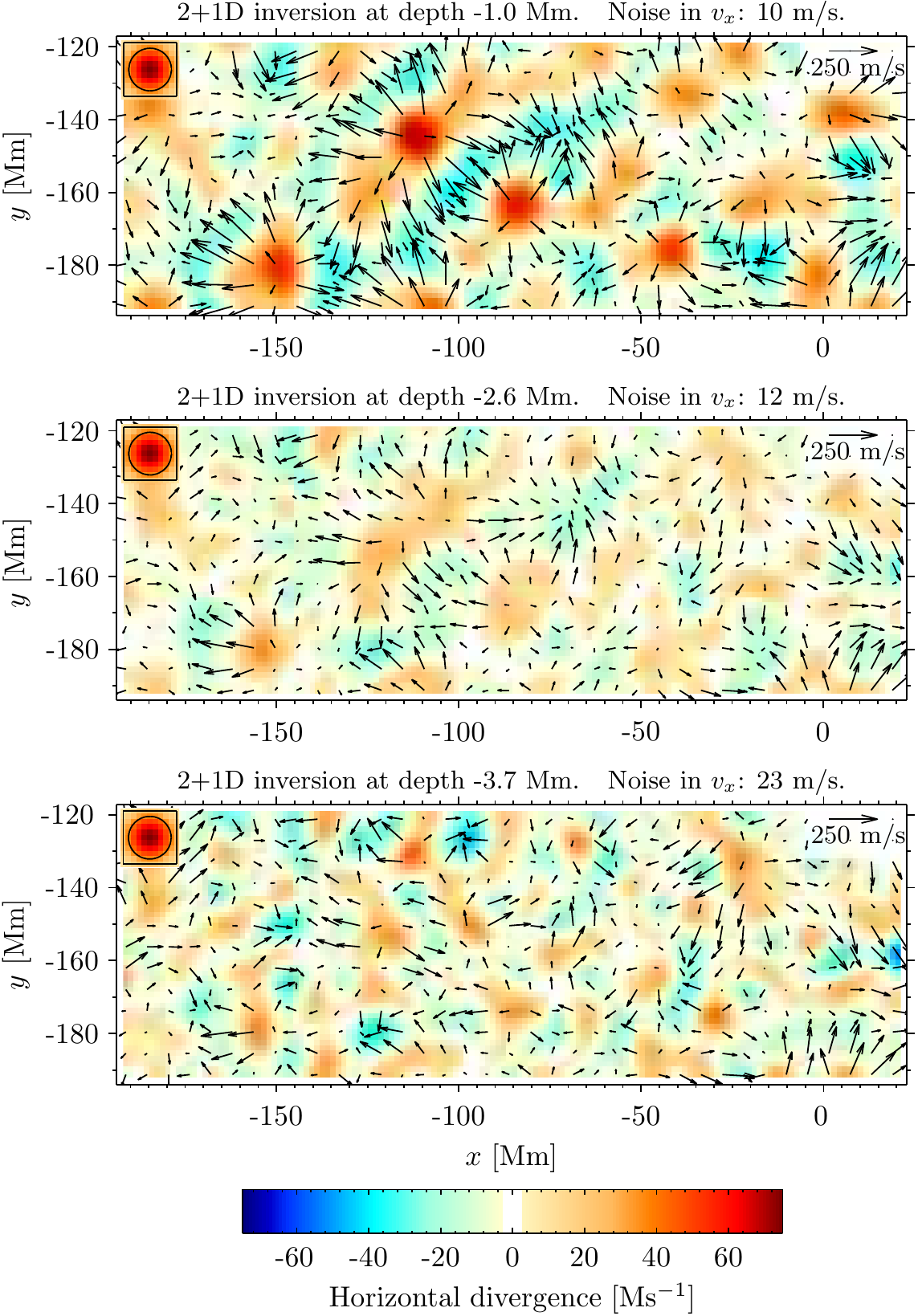}}
  \vspace{.3cm}
  \caption{Comparison of horizontal flows (arrows) at different depths from a 2+1D inversion. The background colorscale is the horizontal divergence to emphasize the various flow structures and  obtained by numerical differentiation of the 2D flows. The $x$ component of the 2D averaging kernels are shown  in the upper left and all have FWHM$=11.6$~Mm. The correlation of the $x$ component of the flows at a depth of $-1$~Mm  and $-2.6$~Mm is $0.77$. The correlation of the $x$ component of the flows at a depth of $-1$~Mm  and $-3.7$ maps is $0.33$. The 1D inversion coefficients for each depth inversion are provided in Table~\ref{tab:weights}.}
  \label{fig:3depths}
\end{figure}

\begin{figure}
\centerline{\includegraphics[width=.8\linewidth]{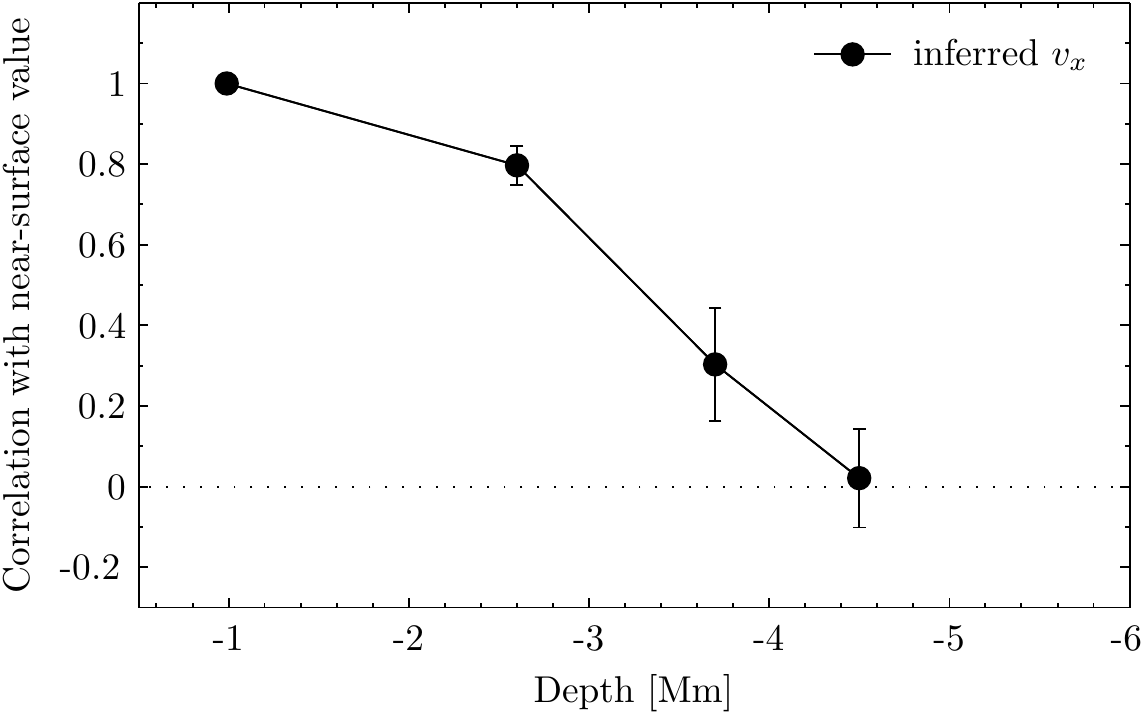}}
 \caption{Correlation with respect to the near-surface value of the inferred $\vv_x$ flows at different depths using $24$~h of data. The values represent the average over $5$~days of the flows measured in the area of the quiet Sun used throughout this paper. The $1\sigma$ error bars are plotted at each depth point, obtained from the standard deviation in the correlations over the $5$~days. The `near-surface value' for this case is at a depth of  $-1$~Mm.}
\label{fig:holo}
\end{figure}

A consistency test of the full 2+1D inversion is shown in Figure~\ref{fig:mode_comp2}. We study flows in a quiet-Sun region  obtained from two independent
inversions using $24$~h of data. Shown are the  horizontal flows computed from inversions for $\vv_x$ and $\vv_y$ along with the line-of-sight magnetic field (colorscale). In the inversion corresponding to panel~(a), we only invert \pone\ travel times. We then attemp to  
target the \pone\ averaging kernel by using  all of the other available 
mode-ridge kernels  \emph{except} \pone, i.e., f, \ptwo, \pthree, and
\pfour. Once  a similar averaging kernel is found, we invert the corresponding f, \ptwo, \pthree, and
\pfour\ travel times, and the resulting flows are shown in Figure~\ref{fig:mode_comp2}b.  The 1D averaging kernels for each case are given in the panel on the bottom
right. The maps are quite similar (correlation  $0.82$), and the differences
could be due simply to the differences in the averaging kernels.   In these maps the supergranule-scale flows are evident, and the
magnetic field is concentrated at the boundaries of the
supergranules as expected. A best fit
through a scatterplot of the  data taking into account the noise in both
variables gives a slope of $0.83$. The magnetic field does not  introduce any anomalous component in the scatter as it did for the full-map study in Figure~\ref{fig:flows_los}, confirming that this is a quiet-Sun region for our purposes. We have also used this test to see if  we can recover an f-mode map by inverting the four available acoustic-mode travel-time sets. For as closely as we are able to match the averaging kernels, it is successful. The same conclusion can be drawn for the other possible cases when noise is not a limiting factor.

In all of the plots studied so far, we have shown flows obtained from $1$~day of travel times.  In Figure~\ref{fig:timeave}, we compare horizontal flows at a depth of $1$~Mm below the surface from inversions for different observation times. The panels show inversions for $6$~h to $24$~h in six hour intervals. Also shown is the horizontal divergence computed numerically (we have not yet computed inversions directly for horizontal divergence at depth). What is evident is that even with $6$~h of data and a resolution of about $9$~Mm, the noise level is reasonable and features are seen that have much in common with the $1$~day map. The correlation between the $6$~h and $24$~h maps is still about $0.8$. It is encouraging that the supergranulation signal at this depth is not dominated by noise for $6$~h of data.

We now compare horizontal flows at three different depths from the full  2+1D inversion. Figure~\ref{fig:3depths} shows the flow field at depths of $1$~Mm (top), $2.6$~Mm (middle), and $3.7$~Mm (bottom) below the surface using 24~h of data. The color scale is the horizontal divergence computed by numerical differentiation of $v_x$ and $v_y$. The 1D inversion weights for each map are given in Table~\ref{tab:weights}. The flows at the different depths in  Figure~\ref{fig:3depths} are not too unlike the individual ridge flows shown in  Figure~\ref{fig:3modes}, and inspection of the 1D inversion weights confirms that this should be the case. This figure also demonstrates that combining the maps with the 1D inversion not only gives a good estimate of the target depth, but also generally lowers the noise levels.

We have studied the correlation of maps of $\vv_x$ and $\vv_y$ such as those in Figure~\ref{fig:3depths} at different depths with the near-surface map and averaged over $5$~days. A plot of the results is provided in  Figure~\ref{fig:holo}. Each measurement is for $24$~h of data, and the error bars are obtained by studying the variance in the correlation values. The correlation steadily decreases as we go deeper, and seems to disappear at about  $5$~Mm below the surface. However, the noise levels at these depths are quite large and we can draw no other specific conclusions at this time. This is consistent with recent studies on realistic numerical simulations  using time-distance helioseismology \cite{zhao2007} and helioseismic holography \cite{braun2007}. In fact, the authors in \inlinecite{braun2007} note that ``\dots supergranule-sized flows are essentially undetectable using current methods below depths around $5$~Mm \dots'' using $24$~h of data or less. We confirm this conclusion here, and note that similar results have also been found with direct modeling techniques \cite{woodard2007}.

We have also studied the  day-to-day correlation of the $\vv_x$ and  $\vv_y$ maps at various depths. If we were predominantly measuring noise, there would be no significant correlation from one $24$~h period to the next. Computing an average day-to-day correlation over seven days of data for the  $-1$~Mm depth maps gives a value of about 0.4.   For the $-3.7$~Mm depth, we find a  0.26 correlation, and at a depth of  $-6$~Mm, about a 0.1 correlation. This again demonstrates that there is plenty of near-surface flow signal when $24$~h averages are studied, presumably due to supergranulation \cite{gizon2004}, which then quickly decreases with depth.

\subsection{Vertical flows}

\begin{figure}
\centerline{\includegraphics[width=\linewidth]{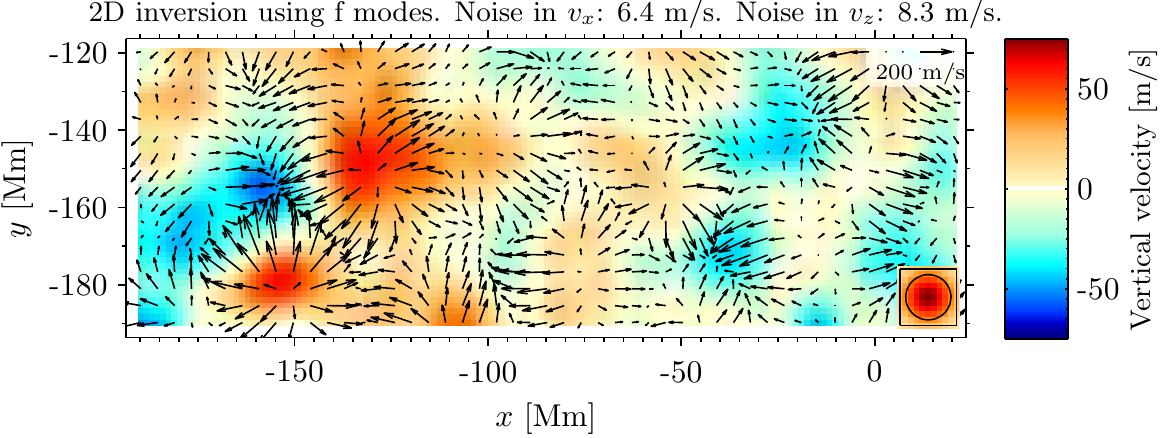}}
\centerline{\includegraphics[width=\linewidth]{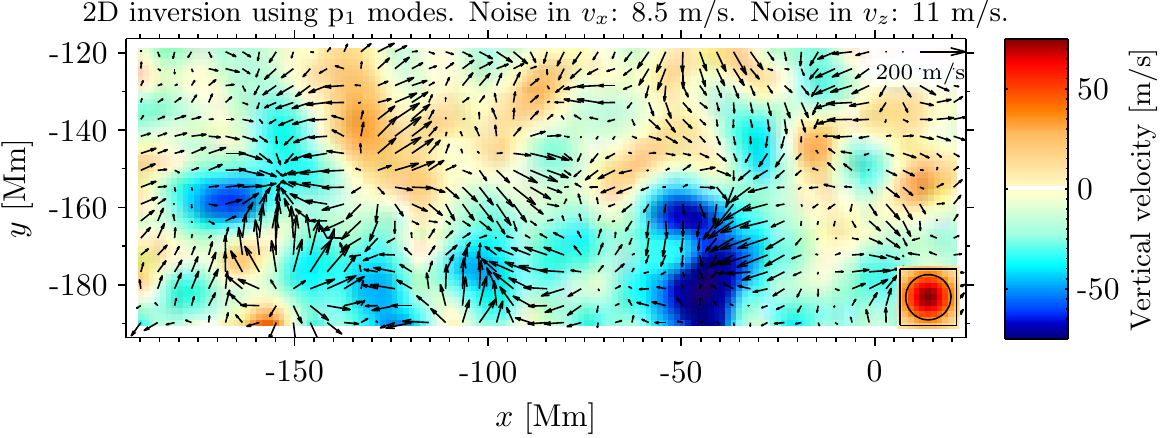}}
\centerline{\includegraphics[width=\linewidth]{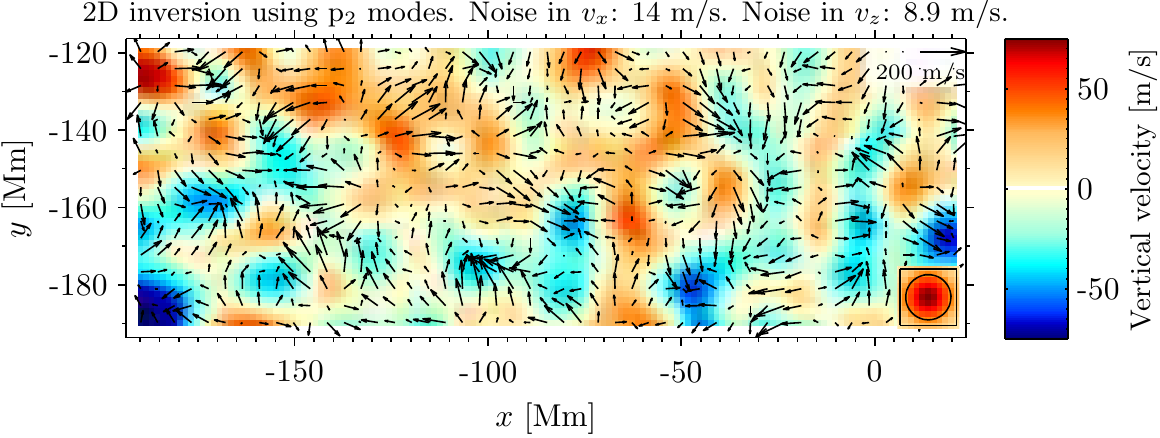}}
\caption{Vertical flows (color scale) and horizontal flows (arrows) in the quiet Sun for a 2D
  inversion using f modes (top), \pone\ modes (middle), and \ptwo\ modes (bottom). These flows were obtained using  $4$8~h of data. A positive vertical velocity means an upflow. The $z$ component of the 2D averaging kernel from the inversion for $\uu_z$ is
  given by the quantity in the box in the lower right. The noise for all of the  measurements is indicated. The correlation of these particular \pone\ and \ptwo\ vertical flows with the f-mode map  is about $0.3$ and $0.2$, respectively. The region of the Sun used here is outlined by the white box in Figure~\ref{fig:flows_los}. The color scale is the same in each panel to ease comparison.}
\label{fig:flows_vz_p2}
\end{figure}

It has proven difficult in helioseismology to accurately measure the vertical component of the velocity near the surface due in part to its small magnitude compared to the other components. In fact, in many helioseismic inversions for flows, $\vv_z$ is
approximately obtained by computing the horizontal flow component and then invoking mass
conservation from the continuity equation (see \opencite{komm2004} for an example in ring-diagram analysis).  
Another source of difficulty in these measurements has been associated with cross-talk effects, whereby the inversion (or sensitivity kernel) becomes insensitive to differences between upflows and convergence, and downflows and divergence \cite{zhao2007}. These inversions, usually based on the ray approximation,  have no obvious means of constraining the cross talk.  Since we have available Born sensitivity kernels for
$\vv_z$, and an inversion procedure which measures each flow component while  minimizing the cross talk with the others, we can
obtain  vertical flows directly and with the assurance that they are relatively independent from the horizontal measurements. This is clearly demonstrated in the averaging kernels of Figures~\ref{fig:mode_avekerns_f_vz} and \ref{fig:mode_avekerns_p1_vz}. We note that we have so far only tested the 2D inversion for $\vv_z$; thus, the maps shown here are for individual ridge measurements.

There tends to be much  more relative noise in the measurements of vertical velocity, and therefore
in Figure~\ref{fig:flows_vz_p2} we show the vertical component of the velocity as the color scale averaged
over 2 days from a 2D inversion (the noise goes as $T^{-1/2}$, where $T$ is the observation time). The top panel of Figure~\ref{fig:flows_vz_p2} is for the f-mode ridge, the middle panel for \pone,  and the bottom panel for \ptwo.  Also shown are the corresponding horizontal flows given by the
arrows.   One generally sees a good correspondence in all maps between the vertical upflows and
horizontal outflows, as well as between downflows and horizontal inflows. Analysis of many similar maps  show that the speeds of the vertical flows in the center of supergranules near the surface are on average about $15-20$~\% of the speeds of the horizontal outflow in the supergranules, slightly higher than recent observations might suggest \cite{hathaway2002}.

\begin{figure}
  \centerline{\includegraphics[width=.8\linewidth]{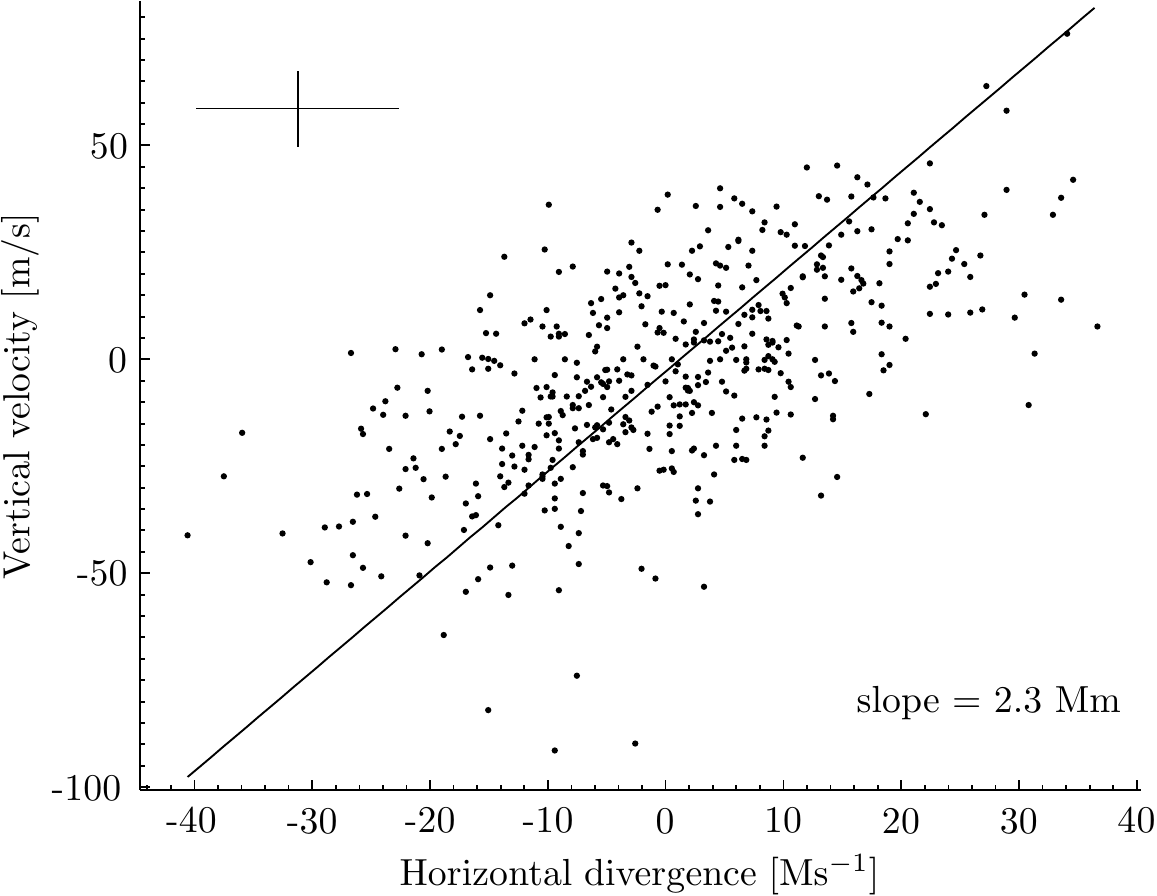}}
\caption{Vertical velocity versus horizontal divergence for flows using  \ptwo\ modes  measured
  over $2$~days for the same region of the Sun as shown in Figure~\ref{fig:flows_vz_p2}. The vertical flows and the horizontal divergence were computed from two separate inversions. The
  correlation coefficient is $0.62$. The line shows a best fit through the scatter taking into
  account the noise on each axis, which is $9$~m\,s$^{-1}$ for $v_z$ and $9$~Ms$^{-1}$ for the divergence, indicated by the cross in the upper left of the figure. The slope of the best-fit line is a rough proxy for the density scale height at the implied depth.}
\label{fig:scatt_vz_p2}
\end{figure}

To understand if the inferred vertical flows at these depths for \ptwo\ are reasonable, we compare them  with maps of the horizontal divergence, obtained from a separate and independent 
inversion of the this quantity as explained in Section~\ref{sec:tests2d}. The vertical component of the flow  and horizontal divergence  are  proportional if one writes down an approximate continuity equation whereby one neglects the horizontal variations in the density and the vertical gradient of the vertical flow. The scaling factor is the density scale height. In Figure~\ref{fig:scatt_vz_p2} we show a scatterplot of $\vv_z^{{\rm p}_2}$ against the  
horizontal divergence inferred from inverting \ptwo\ travel times  for the same region
of the Sun as in Figure~\ref{fig:flows_vz_p2}.  The correlation coefficient is 0.62. The noise in $\vv_z$ is
$9$~m\,s$^{-1}$ and $9$~Ms$^{-1}$ for the divergence measurement. The slope of the best fit line, using the noise information in both variables, gives a value of about $2.3$~Mm. This value is in the range of the density scale height for the implied depth range of these vertical flows. We have also studied the correlation of vertical flows maps with horizontal divergence maps for the f-mode  and \pone-mode cases. The values are always in the range of  0.6-0.7. 

Another  interesting
question is how well the the near-surface vertical flows  are correlated
with deeper vertical flows. Since we have so far only implemented the  2D inversion scheme for vertical flows, we take different mode ridges as a proxy for depth. We correlate the $24$~h f-mode $\vv_z$ map with the \pone\ and \ptwo\ maps, average over seven days,  and find correlations of about 0.3 and 0.2, respectively. In addition, as was described previously for the horizontal component, we  have also studied the day-to-day correlations of the vertical flows averaged over seven days of data. The average day-to-day f-mode map correlation is 0.15, 0.2 for  $\vv_z^{{\rm p}_1}$, and 0.15 for  $\vv_z^{{\rm p}_2}$. This demonstrates  again that the $\vv_z^n$ inversions are measuring long-lived flow structures and not just noise. 

\begin{figure}
\centerline{\includegraphics[width=\linewidth]{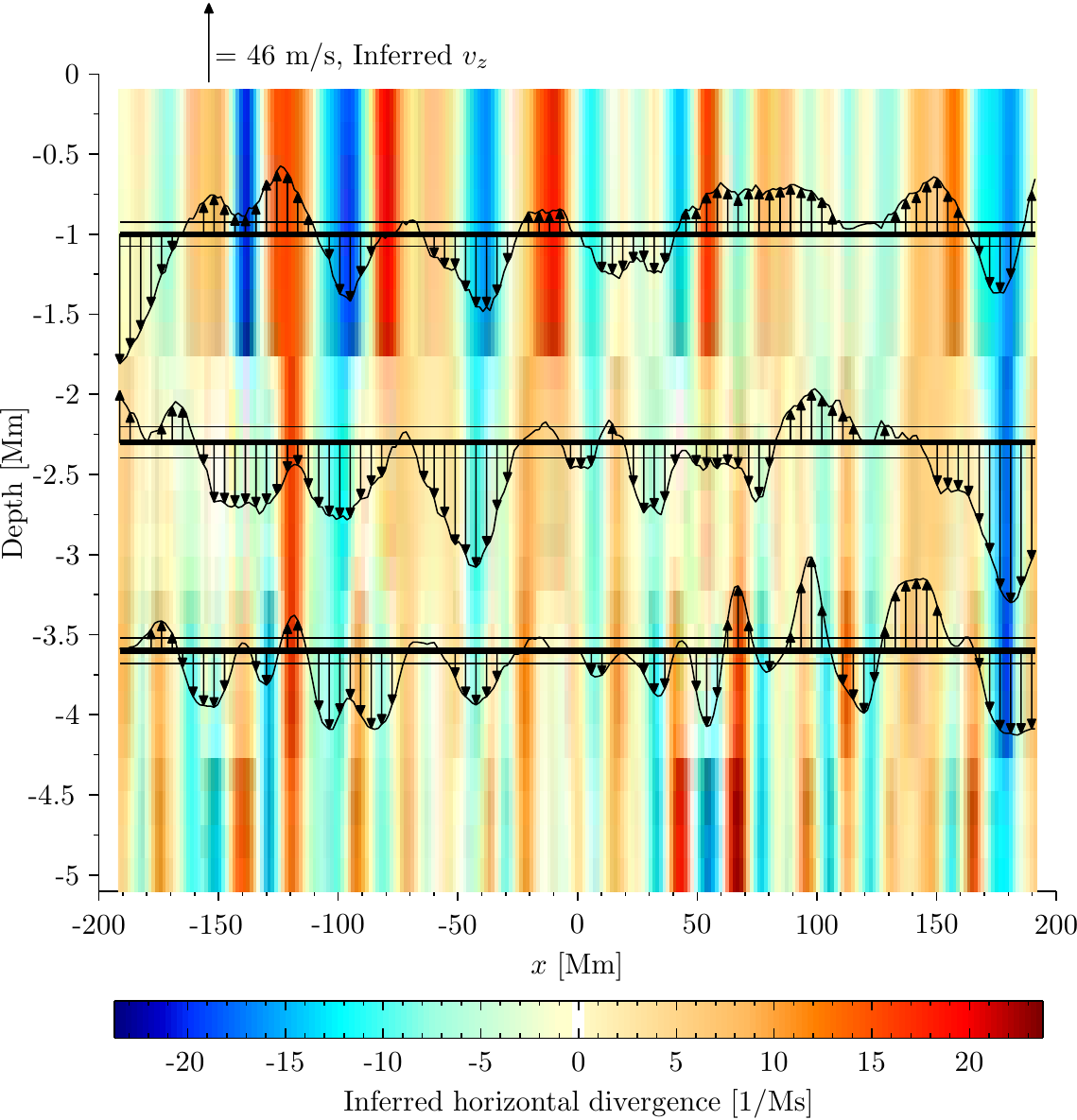}}
\vspace{.5cm}
\caption{Depth slice through a flow map from inversions using 3 days of data.  The horizontal divergence of $\vv_x$  is shown by the color scale and the arrows represent the velocity of the vertical flows at three different depths. The three thick black horizontal lines denote the approximate depth locations of the $\vv_z$ measurements, which are found  by computing the center of mass of the 1D sensitivity functions for the first three ridges (see Fig.~\ref{fig:oned_kerns}). The top curve shows $\vv_z^{\rm f}$, the middle $\vv_z^{{\rm p}_1}$, and the bottom  $\vv_z^{{\rm p}_2}$.  The divergence map was obtained  by computing maps of $\vv_x$ at several different depths where each depth block, centered about the target depth, has a horizontal spatial resolution of $<10$~Mm. Then the numerical divergence was calculated. A positive divergence means an outflow.  The  $\pm 1\sigma$ noise for the three $\vv_z$ measurements are given by the two thin lines above and below the depth indication line, and all have values less than $10$~m\,s$^{-1}$. Note at the top the reference scale arrow for  $\vv_z$.}
\label{fig:depthslice}
\end{figure}

Finally, in Figure~\ref{fig:depthslice} we show the culmination of our main results. It is a slice in depth through the horizontal divergence with overplotted $\vv_z$ information. The inversion to obtain these flows used  $72$~h of travel times. The slice is  along a line through the quiet Sun (shown by the dashed white line in the magnetogram of Figure~\ref{fig:flows_los}) chosen because of the presence of   many near-surface large-scale flow  structures. Inversions for $\vv_x$ at different target depths were performed, the numerical divergence $\partial_x\vv_x$ was computed,  and the results are given by the colorscale. The color scale is such that a positive divergence means an outflow. The $\vv_z$ flows were obtained using f, \pone, and \ptwo\ travel times, and the magnitudes and directions are shown by the arrows. Since we do not yet implement a 1D depth inversion for $\vv_z$, we roughly determine the three depth locations  by computing the average depth over which the dominant modes of these three ridges probe. We emphasize that these placements are only  approximate.  We also plot the $1\sigma$ noise levels of $\vv_z$ at each depth.   We see that over the whole depth range, the horizontal inflows (outflows) generally correspond to vertical downflows (upflows). What one would expect is to see  relatively stronger vertical flows where the divergence is strongest in absolute value. This is for the most part  the case. We emphasize that these good correlations are likely not due to  cross-talk contamination, which tends to diminish as one moves further below the surface (see Section~\ref{sec:2davekerns}). Note also the presence of  large-scale structures that live for at least 3 days.


%


%

\section{Summary and conclusions}

We have presented in detail a fully-consistent procedure for inverting helioseismic travel times  to infer vector flows in the upper convection zone of the quiet Sun. Travel-time sets are measured for all modes that have the same radial order, i.e., along the ridges in the power spectrum. The travel times are constructed using an analogue to the common point-to-annulus geometry for 20 annulus radii (up to about $30$~Mm).  Three-dimensional Born sensitivity kernels for the same travel-time definition and ridge filtering are computed. In addition, the noise covariance properties of the travel times are calculated. Based on the separability of the sensitivity kernels into horizontal and vertical components due to the ridge filtering, the inversion is formulated in two steps: the first step solves the 2D horizontal problem and the second step solves the 1D depth inversion.  Optimal sets of weights are chosen from both inversions, such that the final averaging kernel is regularized in the horizontal and vertical directions. We have provided many examples of averaging kernels, which are extremely useful for understanding where in space is the sensitivity of the inversion, as well as for determining the amount of cross talk among all of the flow components. It was found that the cross talk is reasonably small because the  inversion procedure attempts to minimize its effect by the use of certain constraints. We furthermore obtain consistent estimates of the noise on the measured velocities and the spatial resolution. For practical reasons, the inversion technique is convenient since directly inverting  for other quantities such as the horizontal divergence of the flows or the vertical vorticity only requires one to change the target function.

Many high-resolution example flow maps have been studied and tested. We have restricted ourselves to quiet Sun only. These maps all have horizontal spatial resolution less than about $11$~Mm, or about the \pone-mode wavelength at $3$~mHz.  The recovered flow speeds are below the limits for which a linearized theory of travel times is valid \cite{jackiewicz2007b}. We have tested the inversion in several straightforward ways.  We have shown that using independent measurements and  similar  averaging kernels gives consistent results. We have also been able to obtain high correlations ($\sim 0.9$) with the Doppler velocity data after projecting the inferred horizontal flows onto the line-of-sight vector and ignoring pixels with strong magnetic fields. 

 We also found that the correlation of $24$~h of inferred horizontal flows from day to day on average is about 0.4 near the surface and about 0.1 down to about $6$~Mm below the surface. This is consistent with the conclusion that we are not just measuring noise. However, we find that the correlation of flows at a particular depth with the surface flows falls off quite rapidly and disappears near $\sim 5$~Mm beneath the surface, where we do not see any more evidence of supergranulation.  Similar results have also recently been found on numerical simulations using time-distance helioseismology \cite{zhao2007} and  holography  \cite{braun2007}.  It could be that for $24$~h and at these depths the supergranulation signal is completely masked by noise \cite{braun2007,woodard2007}.

We have  shown a direct inversion for the vertical component of the velocity using acoustic and surface-gravity waves. The results are in agreement with the overall behavior of the horizontal flows, and since the cross talk between $\vv_x$, $\vv_y$,  and $\vv_z$ has been made small, we are fairly confident that the vertical flows are real. The vertical flows have also been compared to independent inversions for the horizontal divergence, and the values  are in the expected ranges. We find that the upflow speeds in the center of supergranules are approximately $15-20$\% of the horizontal outflow speeds. Studying the day-to-day correlations of vertical flow maps also leads us to believe that the signal is above the noise.

Another way to validate many of the findings that we have reported would be to invert the  available artificial velocity data from  realistic numerical simulations of solar convection \cite{benson2006}. Even though the averaging kernels give a complete picture  of how the data is spatially averaged -- a nice feature of OLA-type inversions -- we intend to carry this out in the near future to study the role that noise plays in the interpretation.

Of course, we are undertaking many improvements to the inversion presented here.  One obvious deficiency is the small set of modes we have used. Such a limited number does not allow us to obtain many independent target depths, nor any substantially  deep ones. More ridges, combined with utilizing the spatial frequency content of the waves in each ridge in a more sophisticated way, would help us to obtain better, and deeper, averaging kernels.

Several other improvements currently being studied are ways to  minimize  the cross talk among flow components as much as possible by constructing different types of constraints in the inversion procedure. Also, kernels which take into account the line-of-sight projection are almost certain to be necessary for inverting data well away from disk center. We already have some of these kernels  available  \cite{jackiewicz2006}.

%

%


%



\renewcommand{\theequation}{A\arabic{equation}}


\setcounter{equation}{0}         

\renewcommand{\thefigure}{A\arabic{figure}} 


\setcounter{figure}{0}           

\renewcommand{\thetable}{A\arabic{table}} 


\setcounter{table}{0}            



\bigskip

We thank T. Duvall Jr. for helpful discussions and for providing the data set used in the analysis. We also gratefully acknowledge critical comments from a referee that  significantly improved this paper. \emph{SOHO} is a collaboration between NASA and ESA.




\end{document}